\documentclass[aip,bmf,reprint,superscriptaddress]{revtex4-1}

		\usepackage{amsmath}
		\usepackage{amssymb}
		\usepackage{amsthm}
		\usepackage{mathrsfs}
		\usepackage{esvect}
		\usepackage{latexsym}
		\usepackage{mathtools}
		\usepackage{graphicx}
		\usepackage{epstopdf}
		\usepackage{float}

		\usepackage{amsfonts}
		\usepackage{textgreek}
		\DeclareMathSymbol{\Ohm}{\mathalpha}{operators}{10}

		\usepackage{hyphenat}
		\catcode`@=11
		\MakeRobust{\vphantom}
		\renewcommand{\sin}[1]{\text{sin}\!\left({#1}\right)}
		
		\renewcommand{\cos}[1]{\text{cos}\!\left({#1}\right)}

		\renewcommand{\arcsin}[1]{\text{arcsin}\!\left({#1}\right)}

		\renewcommand{\tanh}[1]{\text{tanh}\!\left({#1}\right)}

		\newcommand{\sech}[1]{\text{sech}\!\left({#1}\right)}

		\newcommand{\logg}[2]{\text{log}_{#1}\!\left({#2}\right)}
		
		\newcommand{\Real}[1]{\text{Re}\!\left({#1}\right)}
		
		\renewcommand{\Re}[1]{\text{Re}\!\left({#1}\right)}

		\renewcommand{\arg}[1]{\text{arg}\!\left({#1}\right)}
		
		\newcommand{\integral}[4]{\int_{#1}^{#2}{#3}\,\text{d}{#4}}
		
		\usepackage[makeroom]{cancel}
		\renewcommand{\parallel}{{\,\scalebox{0.6}{\raisebox{0.8ex}{\cancel{\vphantom{.}\phantom{\,}}\cancel{\vphantom{.}\phantom{\,}}}}}}
		\renewcommand{\leq}{\leqslant}
		
		\renewcommand{\.}{\,\!}
		\newcommand*{\eq}[1]{\begin{eqnarray}#1\end{eqnarray}}

		\newcommand{\figwidth}{\columnwidth}
		
		\newcommand{\graphwidth}{1.2*\figwidth/sqrt(2)}
		\newcommand{\graphheight}{2*\graphwidth/(1+sqrt(5))}
		\usepackage{pgfplots}
		\pgfplotsset{compat=1.9}
		\usepgfplotslibrary{fillbetween}
		\definecolor{darkgreen}{rgb}{0,0.8,0}
		\definecolor{darkred}{rgb}{0.6,0,0}
		\definecolor{grey50}{rgb}{0.5,0.5,0.5}
		\definecolor{grey80}{rgb}{0.8,0.8,0.8}
		\definecolor{lightblue}{rgb}{0.3,0.3,1}
		\definecolor{orangee}{rgb}{1,0.65,0.39}
		\definecolor{yelloww}{rgb}{0.97,0.86,0.15}

		\makeatletter
		\renewcommand\@biblabel[1]{#1.}
		\makeatother
		\usepackage{xifthen}
		\newcommand{\article}[6]{#1 (#2): \textit{#3}, #4 #5, \mbox{#6}}

		\newcommand{\book}[6]{#1 (#2): \textit{#3}, \ifthenelse{\isempty{#4}}{}{\ifthenelse{\equal{\detokenize{#4}}{\detokenize{2}}}{second}{\ifthenelse{\equal{\detokenize{#4}}{\detokenize{3}}}{third}{\ifthenelse{\equal{\detokenize{#4}}{\detokenize{4}}}{fourth}{\ifthenelse{\equal{\detokenize{#4}}{\detokenize{5}}}{fifth}{\ifthenelse{\equal{\detokenize{#4}}{\detokenize{6}}}{sixth}{\ifthenelse{\equal{\detokenize{#4}}{\detokenize{7}}}{seventh}{\ifthenelse{\equal{\detokenize{#4}}{\detokenize{8}}}{eighth}{\ifthenelse{\equal{\detokenize{#4}}{\detokenize{9}}}{ninth}{#4}}}}}}}} edition, }#5, \mbox{#6}}
		\newcommand{\bookarticle}[7]{#1 (#2): \textit{#3}, in \textit{#4}\ifthenelse{\isempty{#5}}{}{ #5}, #6, \mbox{#7}}
		\newcommand{\patent}[5]{#1 (#2): \textit{#3}, #4 \mbox{patent #5}}

		\newcommand{\Ep}{E_\text{p}}
		\newcommand{\vecEp}{\mathbf{E}_\text{p}}

		\newcommand{\Er}{E_\text{r}}
		\newcommand{\vecEr}{\mathbf{E}_\text{r}}

		\newcommand{\thetamax}{\theta_\text{m}}

		\newcommand{\degC}{$^\circ$C}

\begin{document}

\title{Simultaneous microscopic imaging of thickness and refractive index of thin layers by heterodyne interferometric reflectometry (HiRef)}

\author{Alexander Nahmad-Rohen}
	\email{alexander.nahmad@icat.unam.mx}
	\address{Instituto de Ciencias Aplicadas y Tecnolog\'ia, Universidad Nacional Aut\'onoma de M\'exico, Circuito Exterior S/N, Ciudad Universitaria, Ciudad de M\'exico, 04510, M\'exico}
\author{David Regan}
	\address{School of Physics \& Astronomy, Cardiff University, The Parade, Cardiff, CF24 3AA, United Kingdom}
\author{Paola Borri}
	\address{School of Biosciences, Cardiff University, Museum Avenue, Cardiff, CF10 3AX, United Kingdom}
\author{Wolfgang Langbein}
	\email{langbeinww@cardiff.ac.uk}
	\address{School of Physics \& Astronomy, Cardiff University, The Parade, Cardiff, CF24 3AA, United Kingdom}

\date{\today}

\begin{abstract}
The detection of spatial or temporal variations in very thin samples has important applications in the biological sciences. For example, cellular membranes exhibit changes in lipid composition and order, which in turn modulate their function in space and time. Simultaneous measurement of thickness and refractive index would be one way to observe these variations, yet doing it noninvasively remains an elusive goal. Here we present a microscopic-imaging technique to simultaneously measure the thickness and refractive index of thin layers in a spatially resolved manner using reflectometry. The heterodyne-detected interference between a light field reflected by the sample and a reference field allows measurement of the amplitude and phase of the reflected field and thus determination of the complex reflection coefficient. Comparing the results with the simulated reflection of a thin layer under coherent illumination of high numerical aperture by the microscope objective, the refractive index and thickness of the layer can be determined. We present results on a layer of polyvinylacetate (PVA) with a thickness of approximately 80~nm. These results have a precision better than 10\% in the thickness and better than 1\% in the refractive index and are consistent within error with measurements by quantitative differential interference contrast (qDIC) and literature values. We discuss the significance of these results, and the possibility of performing accurate measurements on nanometric layers. Notably, the shot-noise limit of the technique is below 0.5~nm in thickness and 0.0005 in refractive index for millisecond measurement times.
\end{abstract}

\maketitle

\section{Introduction}\label{sec-intro}

Reflectometry has long been used as a noninvasive optical probe for properties of a sample. Interferometric reflectometry (iRef) is an enhancement of reflectometry which uses the interference between two beams to infer the complex amplitude of the reflected field rather than its intensity. Alternatively, it can employ the interference of the reflected field with itself (either different reflection orders or the reflections at different interfaces).

The interference of reflections at different interfaces of a coated glass capillary filled with a liquid has been used to measure the liquid's refractive index.\cite{ref-BornhopUS7130060B2} The same principle has been used on planar layers to measure crystal etch rate\cite{ref-SteinslandSAA86} and diamond crystal growth;\cite{ref-CatledgeDRM9} in both cases, this has been done on layers with thicknesses on the order of microns. Adsorption of polymer layers as thin as 80~nm has also been studied with this type of technique.\cite{ref-MunchJCSFT86} A commercial setup has been used to study biological samples by looking at the interference between light reflected at the interface between air and the sample, which sits on a substrate consisting of a SiO$_2$ layer on top of a Si layer, and light reflected at the SiO$_2$-Si interface. This setup has been used to noninvasively monitor DNA degradation, DNA-protein binding and single-virus deposition on surfaces (using ebolavirus and Marburg virus, both of which have a minor axis of about 80~nm).\cite{ref-AvciS15} It has also been used to observe antibody-antigen and DNA-DNA interaction.\cite{ref-AvciS15,ref-DaaboulBB26} Its accuracy is about 50~nm, although with the aid of nanoparticle labelling (which makes it invasive) the accuracy improves to about 20~nm. On a much larger scale, there is GPS interferometric reflectometry, which consists of a receiver above ground detecting electromagnetic waves emitted by satellite, containing the interference of the direct path with the reflection from the ground; this has been used to measure soil moisture.\cite{ref-ChewGPSS20}

Interferometric scattering microscopy (iSCAT) uses reflectometry of a planar interface to measure the properties of small objects close to the interface. It has been used to observe virus capsid self-assembly,\cite{ref-GarmannPNAS116} the bending of 200-nm microtubules,\cite{ref-AmosJCSS14} the dynamics of coexisting lipid domains\cite{ref-deWitPNAS112} and differences in the lateral diffusion of labelled lipid molecules in supported lipid bilayers depending on the label (fluorophores and metallic nanoparticles).\cite{ref-ReinaJPD51} Differential iSCAT has been used to measure protein secretions by single cells.\cite{ref-GemeinhardtJVE141}

Ellipsometry uses reflectometry at inclined incidence and exploits the differences in the Fresnel coefficients for the polarisation components parallel and perpendicular to the plane of incidence. By measuring the intensity and polarisation of the light reflected by a thin layer, the complex reflection coefficients for the two polarisations are determined, and thus the thickness and/or refractive index of the layer can be determined. Notably, this technique uses a defined angle of incidence, requiring layers which are laterally homogeneous over a size much larger than the light wavelength employed, typically at least 100~\textmu m. The technique has been used to measure the thicknesses of uniform inorganic\cite{ref-HenckJVSTA10} and organic\cite{ref-StrombergJRNBSAPC67A,ref-HoeoekAC73} layers and long-range non-uniform inorganic layers\cite{ref-LiuAO33} with nanometre resolution, as well as the refractive index of metallic substrates on which nanometric layers of water are adsorbed.\cite{ref-McCrackinJRNBSAPC67A} Biological applications of ellipsometry include the measurement of the thickness of macaque retinal nerve layers,\cite{ref-WeinrebAO108} which was measured to be between about 20~\textmu m and about 215~\textmu m. Spectroscopic ellipsometry, combining measurements at multiple wavelengths, has been employed to determine the thickness and (by studying absorption bands) the chemical composition of silane films several hundred nanometres thick,\cite{ref-FranquetTSF441} as well as to measure the thickness and electric permittivity of nanometric tellurium layers and decananometric polypyrrole layers.\cite{ref-ArwinTSF113}

Despite the high sensitivity offered by these techniques, full advantage of the complex reflected field is typically not taken. Most of the aforementioned measurements are only of the thickness or the refractive index (or some other parameter, such as moisture) of the sample. Ellipsometry \cite{ref-ArwinTSF113} can determine both refractive index and thickness, but it requires laterally homogeneous layers over a large scale and does not have microscopic imaging capabilities.

We present here a heterodyne interferometric reflectometry technique (HiRef) which can measure the thickness and refractive index of a sample simultaneously and with microscopic optical-diffraction-limited lateral spatial resolution. We demonstrate accurate measurements on samples tens of nanometres thick, and we discuss the possibility of improving this to a few nanometres in the future, as well as potential applications.

\section{Heterodyne interferometric reflectometry (HiRef)}

We develop here a theoretical description of HiRef, which will be used for quantitative analysis of the thickness and refractive index of thin samples.

\subsection{Reflection by a layer}\label{sec-s}

\begin{figure}
	\begin{center}
	\includegraphics[width=\columnwidth]{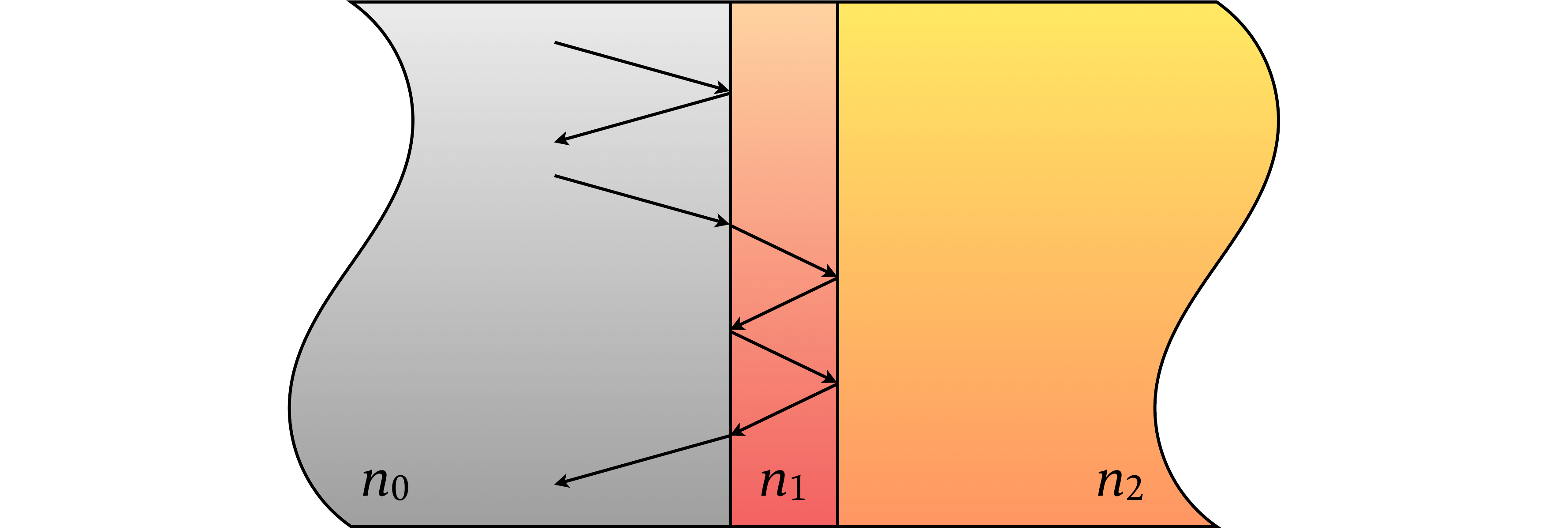}
	\end{center}
	\caption{Sketch of reflection by a homogeneous layer of refractive index $n_1$ between two homogeneous semi-infinite media of refractive index $n_0$ and $n_2$.}
	\label{fig-thinlayer}
\end{figure}

Suppose a homogeneous, isotropic layer of a material with thickness $d$ and refractive index $n_1$ is sandwiched between two homogeneous semi-infinite media with refractive indices $n_0>n_1$ and $n_2<n_1$. Neglecting absorption, a plane light wave incident on the layer (for example, from the $n_0$ side, as shown in figure~\ref{fig-thinlayer}) can be either reflected or transmitted, with the relative amplitude of the reflected and transmitted waves given by the Fresnel coefficients:\cite{ref-HechtO}
\eq{r_{jl}^\parallel & = & \frac{n_j\cos{\theta_j}-n_l\cos{\theta_l}}{n_j\cos{\theta_j}+n_l\cos{\theta_l}},\\
t_{jl}^\parallel & = & \frac{2n_j\cos{\theta_j}}{n_j\cos{\theta_j}+n_l\cos{\theta_l}},\\
r_{jl}^\perp & = & \frac{n_l\cos{\theta_j}-n_j\cos{\theta_l}}{n_l\cos{\theta_j}+n_j\cos{\theta_l}},\\
t_{jl}^\perp & = & \frac{2n_j\cos{\theta_j}}{n_l\cos{\theta_j}+n_j\cos{\theta_l}},}
where the subscript $jl$ indicates that the light is incident on the planar boundary between media with refractive indices $n_j$ and $n_l$ from the $n_j$ side, $\theta_j$ is the angle of incidence, $\theta_l$ is the angle of transmission (given by Snell's law, $n_j\sin{\theta_j}=n_l\sin{\theta_l}$), and the superscripts $\parallel$ and $\perp$ denote the polarisation components parallel and perpendicular, respectively, to the plane of incidence.

If we place a detector on the $n_0$ side, we can collect the light reflected by the layer. The amplitude of the reflected field is given by interference between all the reflection orders. Light incident from the $n_0$ side is reflected at the first boundary with a relative amplitude $s_0=r_{01}$, where the superscript is omitted for clarity in the following calculations. Light transmitted into the second semi-infinite medium is not measured. Light transmitted back into the first semi-infinite medium after having been reflected a total of $j$ times at the $n_1$-$n_2$ interface and $j-1$ times at the $n_0$-$n_1$ interface has the relative complex amplitude
\eq{s_1 & = & t_{01}t_{10}\sum_{j=1}^\infty r_{10}\.^{j-1}r_{12}\.^j e^{ij\Delta},}
where $\Delta=2kdn_1/\cos{\theta_1}$ is the phase due to propagation through the layer and back, $k=2\pi/\lambda$ is the wave number of the light of wavelength $\lambda$, and $\theta_1$ is the propagation angle in the layer. The relative complex amplitude of the detected wave is then
\eq{s & = & s_0+s_1 = r_{01}+t_{01}r_{12}t_{10}e^{i\Delta}\sum_{j=0}^\infty\left(r_{10}r_{12}e^{i\Delta}\right)^j\nonumber\\
& = & r_{01}+\frac{t_{01}r_{12}t_{10}e^{i\Delta}}{1-r_{10}r_{12}e^{i\Delta}}\,,\label{eq-s}}
where $|r_{10}r_{12}e^{i\Delta}|<1$ for $\theta_0$ outside the total internal reflection (TIR) regime. Note that if there is no layer (i.e.~$d=0$) then the right side of equation~\ref{eq-s} reduces to $r_{02}$, the reflection coefficient at the boundary between the two semi-infinite media.

\begin{figure}[b!]
\begin{center}
\begin{minipage}[m]{\columnwidth}
\includegraphics[width=\columnwidth]{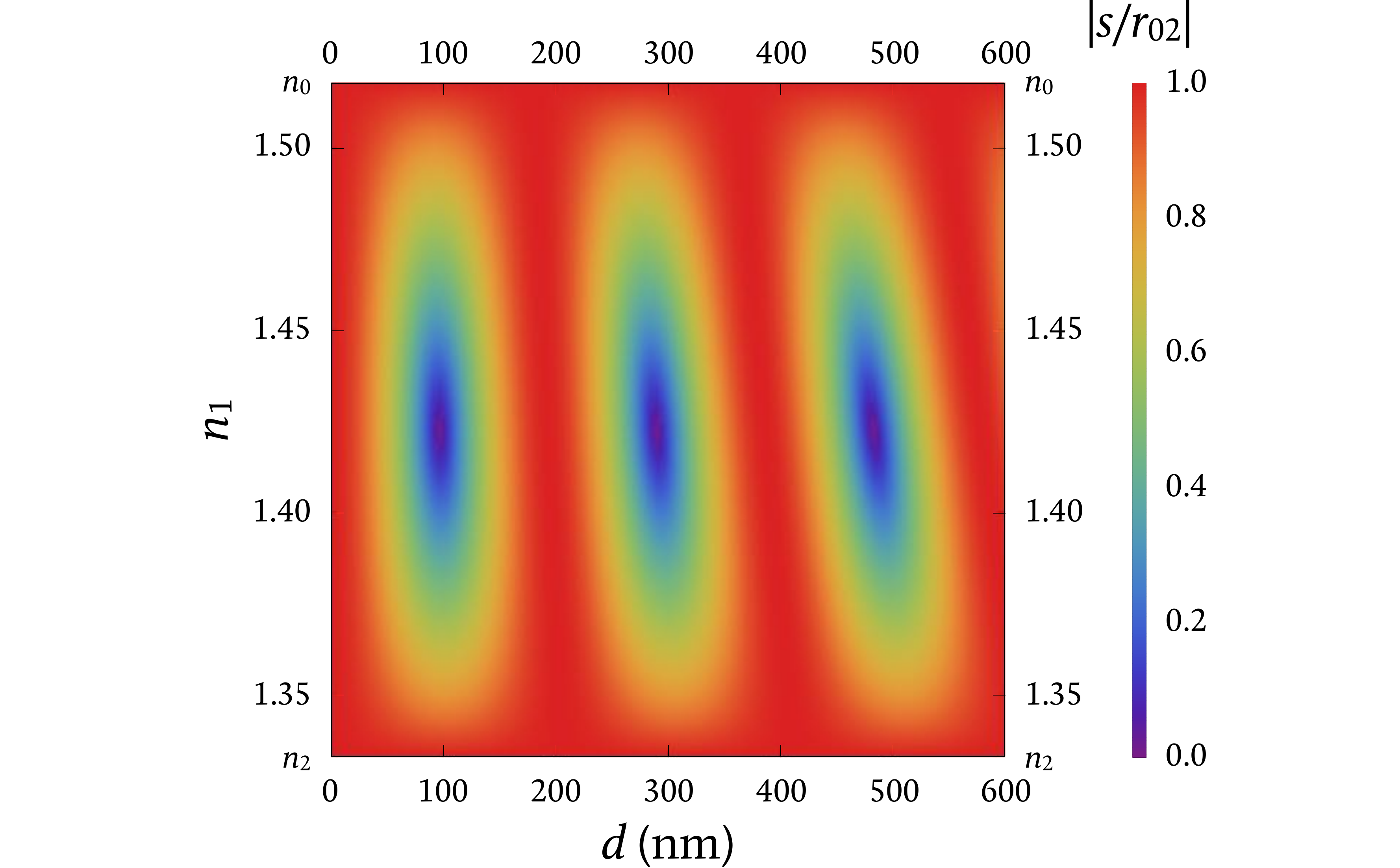}
\end{minipage}\\[2.5ex]
\begin{minipage}[m]{\columnwidth}
\includegraphics[width=\columnwidth]{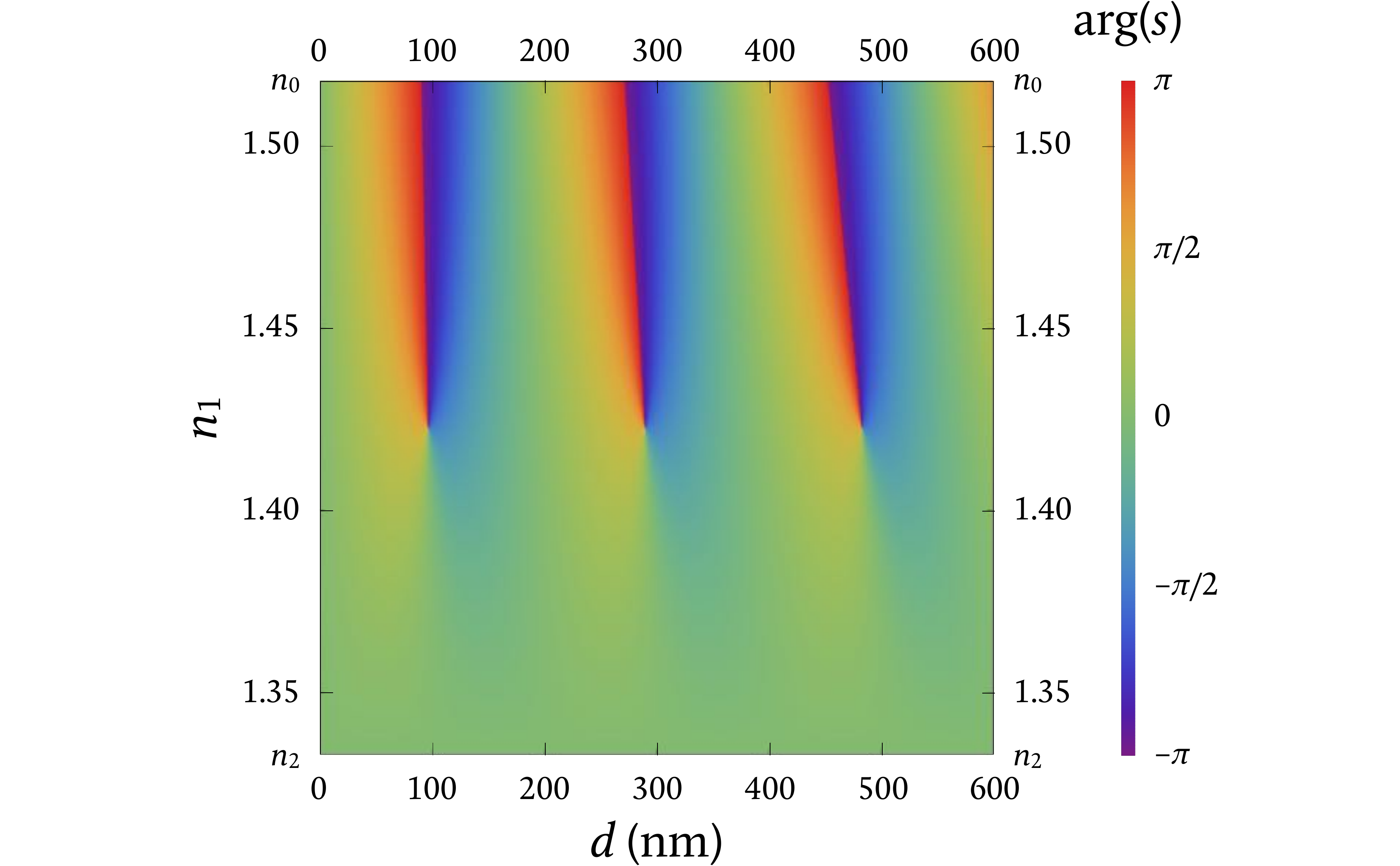}
\end{minipage}
\caption{$|s/r_{02}|$ (top) and $\arg{s}$ (bottom) as functions of $d$ and $n_1$ for normal incidence with $n_2=1.333$, $n_0=1.518$ and $\lambda=550$~nm.}\label{fig-s0}
\end{center}
\end{figure}

For normal incidence ($\theta_0=0$), equation~\ref{eq-s} can be written as
\eq{s & = & \frac{(n_0+n_1)(n_1-n_2)e^{i\Delta}+(n_0-n_1)(n_1+n_2)}{(n_0-n_1)(n_1-n_2)e^{i\Delta}+(n_0+n_1)(n_1+n_2)}\,.\label{eq-snorm}}
Figure~\ref{fig-s0} shows the resulting amplitude and phase of $s$ for $n_0=1.518$, $n_2=1.333$, $\lambda=550$~nm, layer thicknesses between $0$ and $600$~nm and layer refractive indices between $n_2$ and $n_0$. The aforementioned values of $n_0$ and $n_2$ have been chosen to match microscope-slide glass\cite{ref-OcheiMLS} and water\cite{ref-HechtO}, respectively. It is evident that $s$ is periodic in $d$ with a period that is decreasing with $n_1$, as expected from the term $e^{i\Delta}$ in equation~\ref{eq-s}; the thickness range was chosen to show three such periods. It is also clear that $|s|\leq r_{02}$ for any $d$ as long as $n_2\leq n_1\leq n_0$.

\subsection{Interferometric detection of reflection focussed by a microscope objective}\label{sec-HiRef}

HiRef makes use of the heterodyne-detected interference between two beams, one of which (called the probe beam) interacts with the sample and one of which (called the reference beam) acts as an external reference, to obtain the reflected amplitude and phase. This allows the determination of two properties of the sample, such as thickness and refractive index, simultaneously.

We consider here an aplanatic objective of numerical aperture $\text{NA}$. The light field incident on the sample has a wide angular distribution over the polar angle $\theta$, and the maximum angle of incidence is given by $\thetamax = \arcsin{\text{NA}/n_0}$. The probe beam entering the objective back-focal plane is chosen to be circularly polarised, so the probe field in cartesian polarisation basis is given by
\eq{\vecEp(\theta) & = & \frac{E(\theta)}{\sqrt{2}}{\begin{pmatrix}1\\i\end{pmatrix}}e^{-i\omega t}}
with
\eq{E(\theta) & = & E_0\sqrt{\cos{\theta}}\,e^{-\left(\zeta\sin{\theta}/\sin{\thetamax}\vphantom{b^2}\right)^2},}
where we assume a gaussian beam profile in the back-focal plane of the objective with inverse objective fill factor $\zeta$.\cite{ref-NovotnyPONO} The fields parallel and perpendicular to the plane of incidence, which are not mixed by the reflection, are then given by
\eq{\Ep^\parallel(\theta,\varphi) & = & \frac{1}{\sqrt{2}} E(\theta) e^{i(\varphi-\omega t)},\\
\Ep^\perp(\theta,\varphi) & = & \frac{i}{\sqrt{2}}E(\theta) e^{i(\varphi-\omega t)},}
where $\varphi$ is the azimuthal angle. The probe beam is reflected by the sample and recollimated by the objective, and it then interferes with a frequency-shifted reference beam, which does not interact with the sample, in an image plane of the objective back-focal plane. The reference beam is also circularly polarised and has the same beam profile as the probe beam. The reflection inverts the circular polarisation relative to the propagation direction, so we use a reference beam which has opposite circular polarisation to the beam entering the objective. We can therefore assume that its field distribution in the image plane of the back-focal plane of the objective is equal to that of the probe beam, so, up to a constant phase shift, $\vecEr(\theta)=e^{-i\Omega t}\vecEp(\theta)$, where $\Omega$ is the heterodyne frequency shift (see section~\ref{subsec-setup}). We thus have, for each incidence direction $(\theta, \varphi)$, the interference term
\eq{I(\theta,\varphi) & = & 2\Re{\Er^{\parallel\ast}s^\parallel\Ep^\parallel+\Er^{\perp\ast}s^\perp\Ep^{\perp}}\nonumber\\
& = & -|E(\theta)|^2\Re{\!\left(s^\parallel(\theta)+s^\perp(\theta)\right)e^{i\Omega t}\vphantom{\left(e^{i\Omega t}\right)^\frac{b}{b}}\!}\!,}
where the star indicates the complex conjugate. The interference does not depend on $\varphi$ because we use co-circular polarisations for the probe and reference beams. A dual-channel lock-in detection recovers both the in-phase and in-quadrature components at the frequency $\Omega$, so we can detect the complex signal
\eq{S & = & S_0\integral{0}{\thetamax}{|E(\theta)|^2\left(s^\parallel(\theta)+s^\perp(\theta)\right)\sin{\theta}}{\theta},\label{eq-savg}}
where $S_0$ is a constant which depends on the laser powers used, the detection efficiency and other setup-specific constants. The parameters used here are such that TIR does not occur, as the NA of the objective is below the lowest refractive index of the layer structure considered. Therefore, the expressions derived in section \ref{sec-s} for $s$ are valid. Expressions for $s$ including the case of TIR can be derived in a similar fashion.

\begin{figure}
\begin{center}
\begin{minipage}[m]{\columnwidth}
\includegraphics[width=\columnwidth]{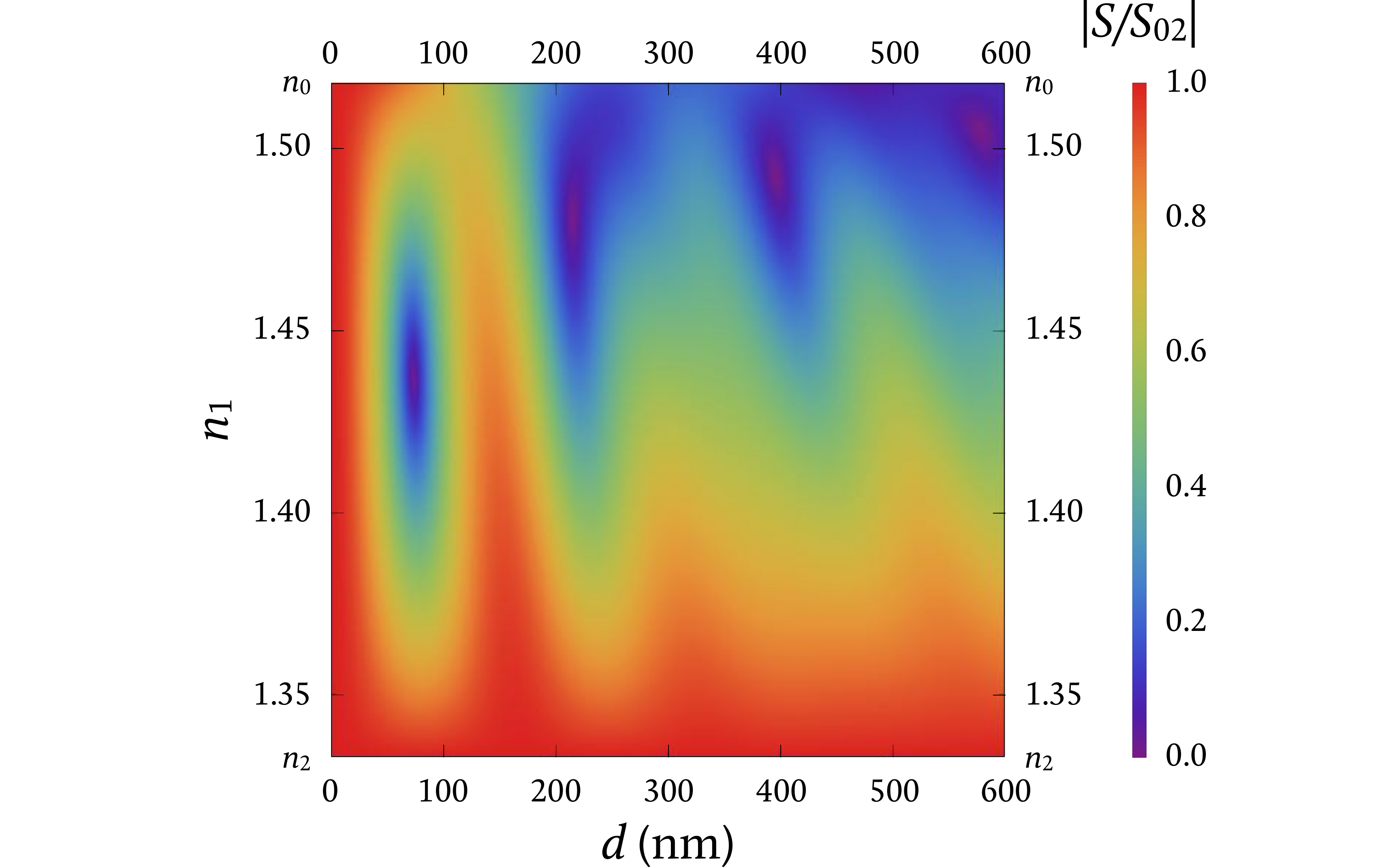}
\end{minipage}\\[2.5ex]
\begin{minipage}[m]{\columnwidth}
\includegraphics[width=\columnwidth]{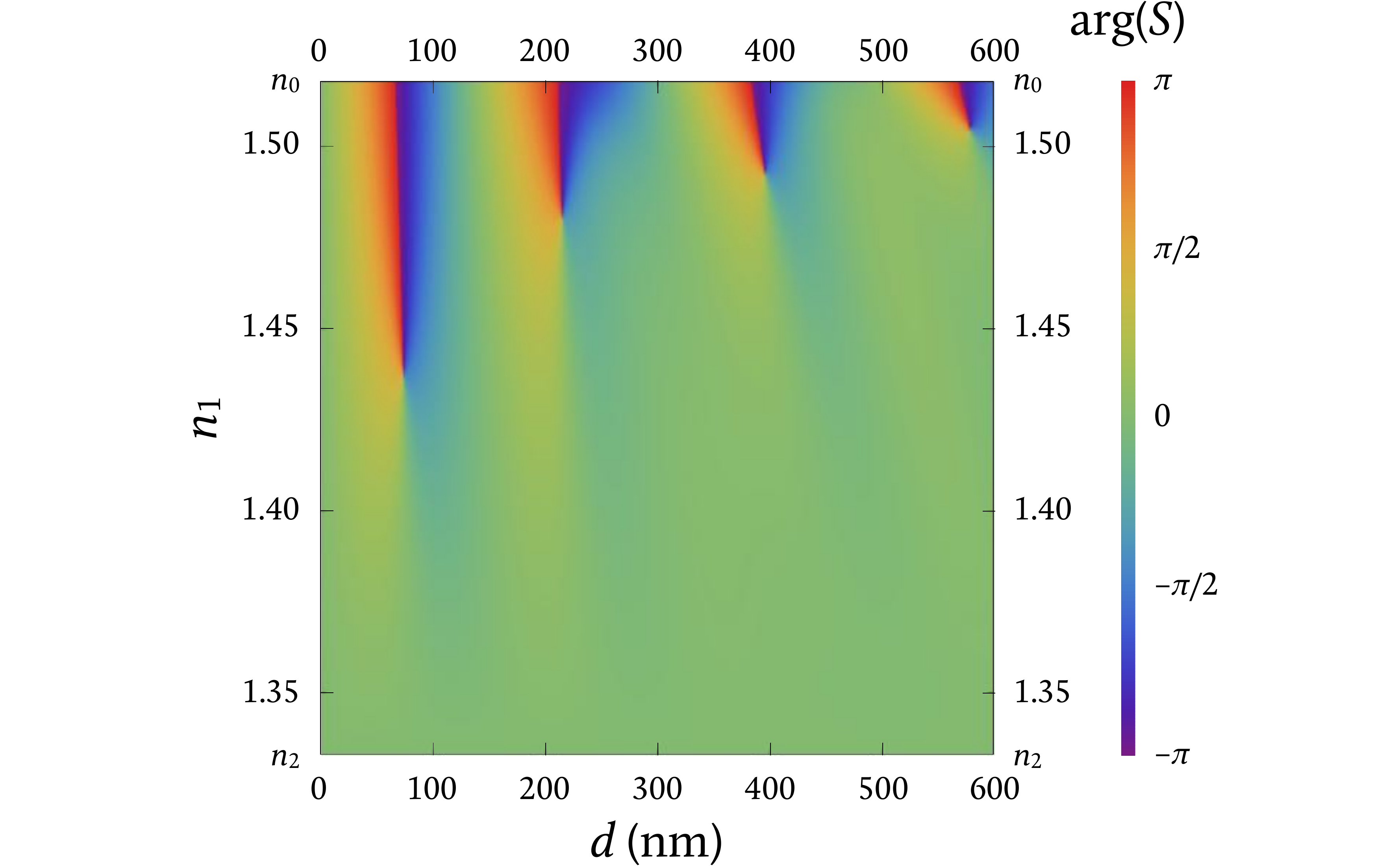}
\end{minipage}
\caption{$|S/S_{02}|$ (top) and $\arg{s}$ (bottom) as functions of $d$ and $n_1$ for an aplanatic objective with numerical aperture 1.27 and fill factor 1, materials with refractive indices $n_2=1.333$ and $n_0=1.518$, and circularly polarised light with wavelength $\lambda=550$~nm.}\label{fig-s127}
\end{center}
\end{figure}

Figure~\ref{fig-s127} shows the amplitude and phase of $S/S_{02}$, where $S_{02}$ is the signal detected in the absence of the thin layer, for which $s^\parallel=r_{02}^\parallel$ and $s^\perp=r_{02}^\perp$. Note that $S_{02}$ is real. The wavelength and refractive indices used are the same as in figure~\ref{fig-s0}; $\thetamax=56.8^\circ$, as determined by $\text{NA}=1.27$; and $\zeta=1$. We find that the qualitative pattern seen with normal incidence remains, but it is distorted, squeezed upwards and towards the left, with increasing distortion as the thickness increases. The signal still satisfies $|S|\leq S_{02}$.

\subsection{Retrieving thickness and refractive index}

\begin{figure}[b!]
\begin{center}
\begin{minipage}[m]{\columnwidth}
\quad\includegraphics[width=\columnwidth]{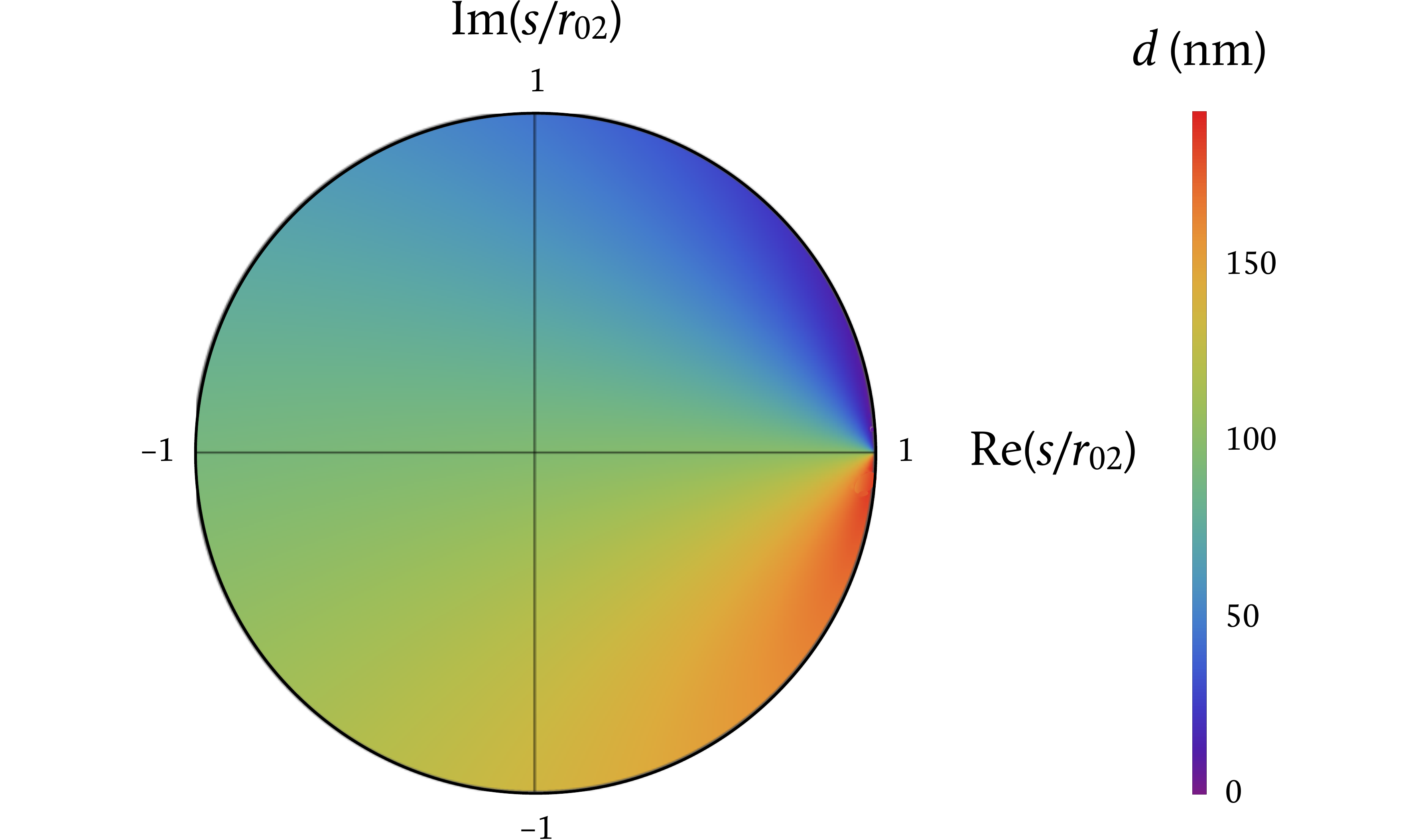}
\end{minipage}\\[2.5ex]
\begin{minipage}[m]{\columnwidth}
\quad\includegraphics[width=\columnwidth]{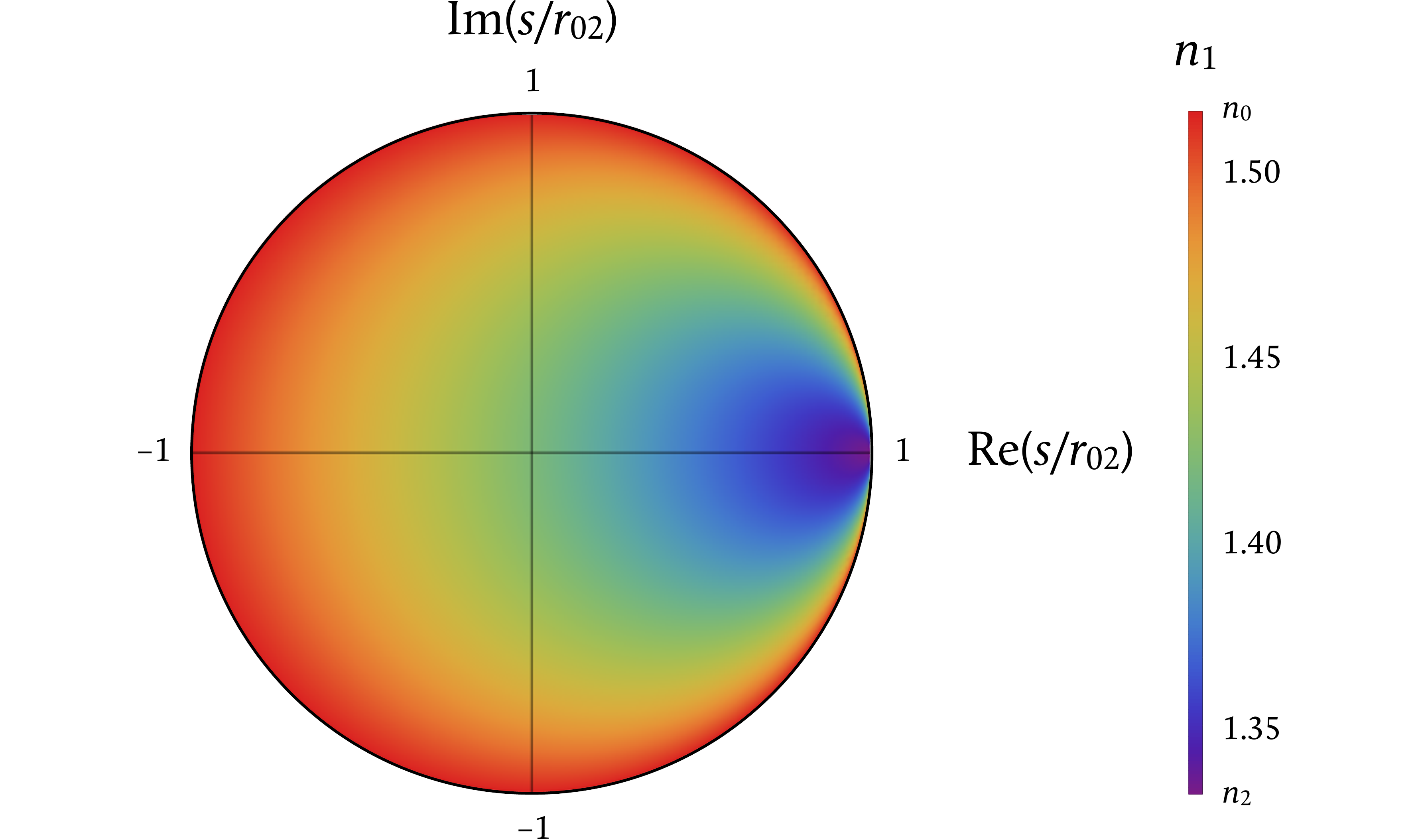}
\end{minipage}
\caption{$d$ (top) and $n_1$ (bottom) as functions of $s/r_{02}$ for normal incidence with $n_2=1.333$, $n_0=1.518$ and $\lambda=550$~nm.}\label{fig-dn0}
\end{center}
\end{figure}

For normal incidence, equation~\ref{eq-s} can be solved analytically for $d$, yielding
\eq{d & = & \frac{i}{2kn_1}\,\logg{}{\frac{n_2-n_1}{n_2+n_1}\,\frac{n_0+n_1-(n_0-n_1)s}{n_0-n_1-(n_0+n_1)s}}.\nonumber\\\label{eq-d}}
Now, since $d$ is a real quantity, the argument of the logarithm in equation~\ref{eq-d} has a magnitude of 1. This condition can be used to obtain
\eq{n_1 & = & \sqrt{n_0 n_2\,\frac{n_0-n_2-2n_0\Real{s}+(n_0+n_2)|s|^2}{n_0-n_2-2n_2\Real{s}-(n_0+n_2)|s|^2}}\,.\nonumber\\}
Figure~\ref{fig-dn0} shows the resulting thickness and refractive index of the layer as a function of $s/r_{02}$, where we have chosen the solution (the branch of the complex logarithm) having the smallest thickness range.

We can see that around the circumference, changing the phase of $s/r_{02}$ while keeping its magnitude close to 1, the thickness of the layer changes while its refractive index remains close to $n_0$. Going instead radially, changing the real part of $s/r_{02}$ starting from 1, the refractive index increases from $n_2$ for $s=1$ to $\sqrt{n_2 n_0}$ for $s=0$ to $n_0$ for $s=-1$.

\begin{figure}
\begin{center}
\begin{minipage}[m]{\columnwidth}
\quad\includegraphics[width=\columnwidth]{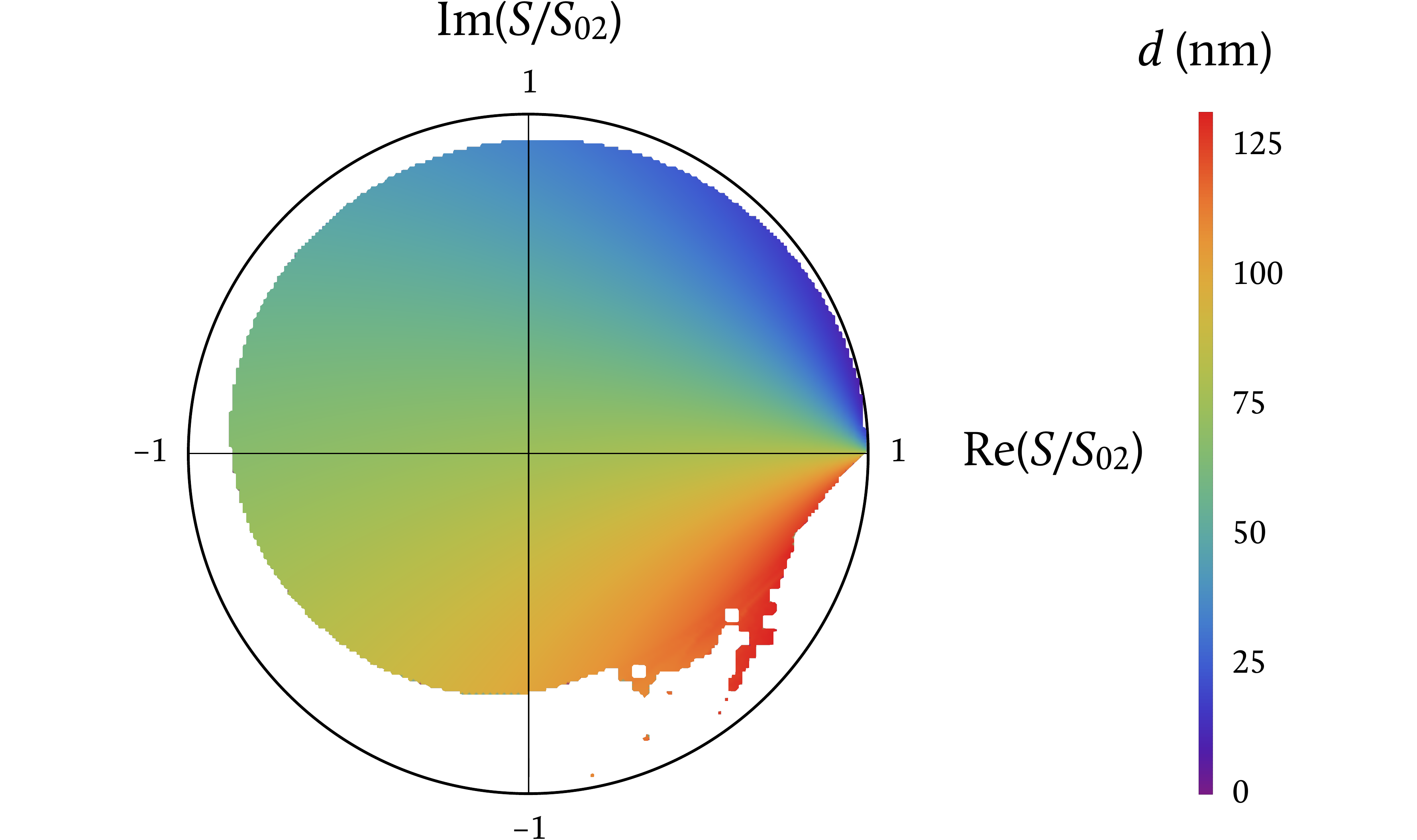}
\end{minipage}\\[2.5ex]
\begin{minipage}[m]{\columnwidth}
\quad\includegraphics[width=\columnwidth]{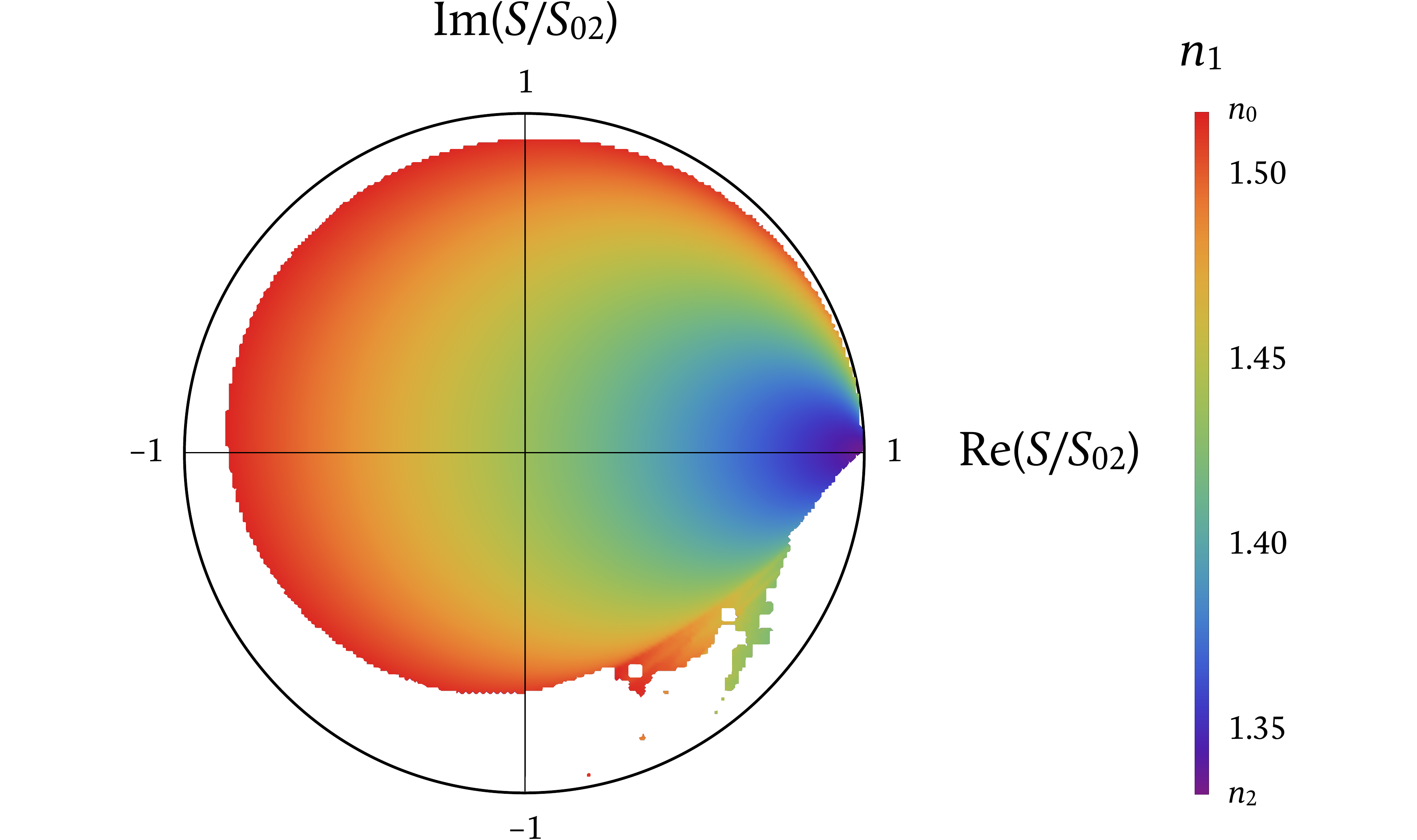}
\end{minipage}
\caption{$d$ (top) and $n_1$ (bottom) as functions of $S/S_{02}$ for an aplanatic objective with numerical aperture 1.27 and fill factor 1, materials with refractive indices $n_2=1.333$ and $n_0=1.518$, and circularly polarised light with wavelength $\lambda=550$~nm.}\label{fig-dn127}
\end{center}
\end{figure}

In the case of reflection of a focussed beam, given by equation~\ref{eq-savg}, and shown in figure~\ref{fig-s127}, the refractive index and thickness of the layer can be determined numerically for the angular distribution we have considered. Figure~\ref{fig-dn127} shows the resulting $d$ and $n_1$ as functions of $S/S_{02}$.

We find that the qualitative behaviour is similar to the one at normal incidence shown in figure~\ref{fig-dn0}, apart from a region with no solutions (white) given by $|S/S_{02}|>R(\arg{S/S_{02}})$. The radius $R$ of this region decreases with increasing angle $\arg{S/S_{02}}$, down to about 0.7 in the fourth quadrant, before increasing back to 1, joining the $S/S_{02}=1$ point of zero thickness. This behaviour reflects the angular averaging of the reflection coefficient, which for a finite thicknesses $d$ reduces $|S/S_{02}|$ below its maximum of 1.

Measuring $S/S_{02}$ therefore allows the determination of the refractive index and the thickness with high spatial resolution given by the sub-micron focus size created by the objective.

\section{Materials \& methods}\label{sec-methods}

\subsection{Experimental setup}\label{subsec-setup}

\begin{figure}[b!]
	\begin{center}
	\includegraphics[width=\columnwidth]{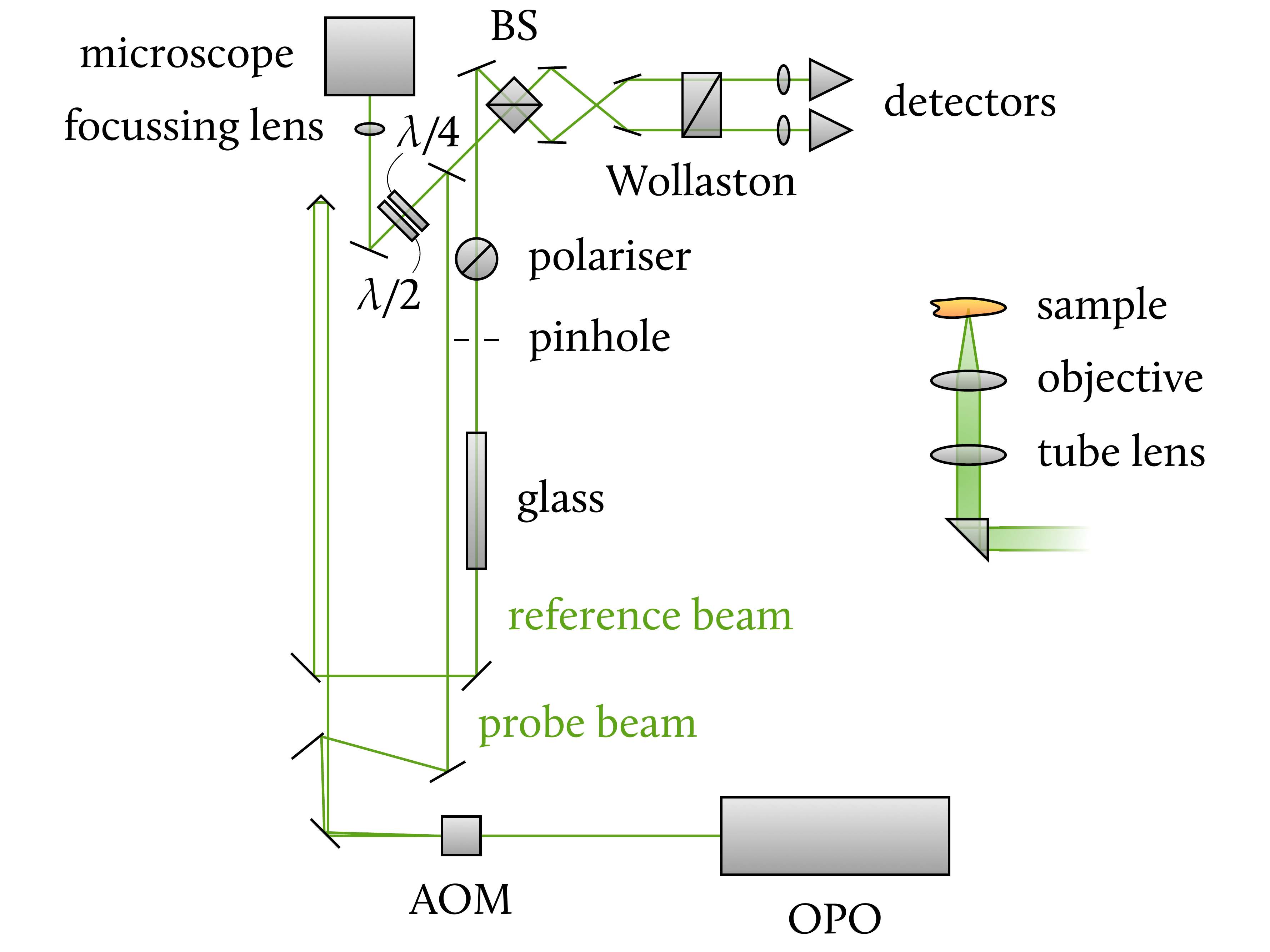}
	\end{center}
	\caption{Schematic diagram of the HiRef setup used. OPO: optical parametric oscillator; BS: beam-splitter; $\lambda/4$: quarter-wave plate; $\lambda/2$: half-wave plate.}
	\label{fig-setup}
\end{figure}

The experimental setup is shown schematically in figure~\ref{fig-setup} and described below, with more detail given elsewhere.\cite{ref-ZorinyantsPRX7} A titanium-sapphire laser (Spectra-Physics Mai Tai) emits 100-fs pulses centred at a wavelength of 820~nm with 80-MHz repetition rate, which pump an optical parametric oscillator (OPO, Inspire Radiantis). The OPO emits a beam of 150-fs pulses centred at 550~nm with horizontal (in the plane of the schematic) linear polarisation. This beam enters an acousto-optic modulator (AOM, IntraAction ASM-802B67) driven at 82~MHz. The zeroth- and first-order deflected beams are used as reference and probe beams, respectively, for HiRef. The probe beam is reflected by a 20R/80T beam-splitter (Edmund Optics NT68-367) and transmitted through a quarter-wave plate (Casix WPA400-550-4) and a half-wave plate (Casix WPA400-550-2), which allow full control of its polarisation state. After the wave plates, the beam is reflected by a dichroic mirror (Eksma Optics, HR415-550nm/HT630-1300nm), transmitted through a 10-mm-thick SF2 glass block used to adjust the beam centring in the back-focal plane of the objective by tip and tilt, and focussed by a lens with a focal length of 100~mm (Edmund Optics T49-360) into the image plane of an inverted microscope (Nikon Ti-U). The beam enters the right port of the microscope, is reflected upwards by the port prism, is collimated by a $1.5\times$ tube lens with a focal length of 300~mm, and is focussed onto the sample by a $60\times$ 1.27-NA water-immersion objective (Nikon MRD70650). The power at the sample is about 10~\textmu W.

The beam reflected by the sample travels back through the optics and is transmitted by the 20R/80T beam-splitter. The beam is then recombined with the reference beam by a low-polarising 45R/45T beam-splitter (BS, Optosigma 039-0235). The reference beam's optical path length is matched to the probe beam's path length by an optical delay line using a linear stage (Physik Instrumente M-403.6DG) and a fused-silica corner cube (Eksma Optics \#340-1217M+3217), and its group velocity dispersion is matched to the probe beam's using 142~mm of SF2 glass (Schott). Note that the probe makes a double pass through the microscope optics described above, accumulating the corresponding amount of dispersion. A linear polariser at 45$^\circ$ creates a well-defined polarisation state of the reference beam with fields of equal amplitude and phase in the horizontal and vertical polarisations.

The two outputs of the BS are separated into horizontal and vertical polarisations by a Wollaston prism and are focussed onto silicon diodes (Hamamatsu S5973-02) to allow balanced detection of the horizontal and vertical polarisation components separately. The reference beam's power was chosen to be around 0.5~mW per diode to ensure shot-noise-limited detection. The differential diode current corresponding to each polarisation is amplified with a transimpedance of 100~k\textOmega\ and analysed by a dual-channel lock-in amplifier (Zurich Instruments HF2) locked to the difference of 2~MHz between the AOM upshift and the pulse repetition rate. The lock-in amplifier extracts the interference between the probe and reference beams in amplitude and phase.

The dual-polarisation detection is used to adjust the polarisation at the sample to be circular on axis as follows: A circular polarisation changes helicity upon reflection on an in-plane-isotropic surface, resulting in a cross-circularly-polarised reflected beam emerging from the sample, which after returning through the wave plates is cross-polarised to the horizontal input polarisation. Therefore, minimising the detected signal in the horizontal polarisation using the wave plates ensures a circular polarisation at the sample. Typically, a horizontal field amplitude below 1\% of the vertical field amplitude can be stably achieved.

The reflected probe beam $E_\text{p}$, with power $|\Ep|^2$, and the reference beam $E_\text{r}$, with power $|\Er|^2$, interfere at the BS, yielding two beams with powers $S_1=|E_\text{p}+E_\text{r}|^2/2$ and $S_2=|E_\text{p}-E_\text{r}|^2/2$. The difference $S_1-S_2=2\Real{\Ep\Er\.^\ast}$ is measured by the balanced detector and analysed by the lock-in amplifier to obtain the amplitude and phase of the combined field. Because the amplitude and phase of $\Er$ are independent of the sample and ideally constant, the amplitude and phase of the reflected probe beam can be determined.

Due to the high NA used, the focal depth of the focussed probe beam is of the order of 1~\textmu m, much smaller than the thickness of the coverslip and the depth of the well formed by the imaging spacer (see section~\ref{subsec-sample}), so the glass and oil layers enclosing the sample layer can be considered infinitely thick for the purpose of interference of reflected beams. Furthermore, the short coherence length of the 100-fs pules (around 30~\textmu m) suppresses interference with reflections occuring at other surfaces.

The sample is moved by an $xyz$ piezoelectric stage with nanometric position accuracy (MadCityLabs NanoLP200) and scanned during image acquisition. Images were acquired with a pixel size $\delta_x=108$~nm at 0.2~ms per pixel. Each image was a square 80~\textmu m on a side and took around 110~s to acquire.

Immediately before each field of view was imaged with HiRef, the same field of view was imaged with quantitative differential interference contrast (qDIC), which provides an accurate measurement of the thickness under the assumption that the refractive index is known.\cite{ref-ReganL35}

\subsection{Sample preparation}\label{subsec-sample}

\begin{figure}[b!]
	\begin{center}
		\includegraphics[width=\columnwidth]{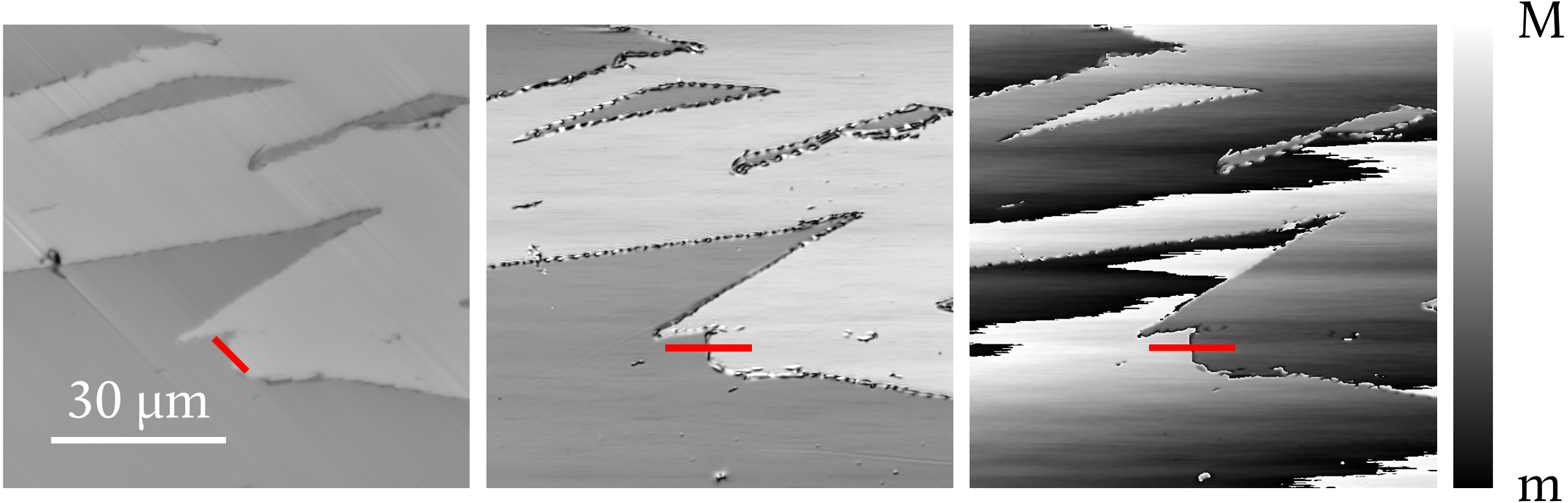}
	\end{center}
	\caption{qDIC phase (left; $m=-0.9212$, $M=0.5581$), HiRef amplitude (centre; $m=9.8$~mV, $M=76.5$~mV) and HiRef phase (right; $m=-\pi$, $M=\pi$) images of one of the studied fields of view of the PVA layer. In the left and centre images, brighter regions are empty and darker regions are the PVA layer. Considerable phase drift is visible in the right image. The red lines show examples of line trace positions (see section~\ref{subsec-analysis}) taken along the shear for qDIC and along the fast scan axis for HiRef. Note that the trace is centred on the same point at a step in all three images.}
	\label{fig-PVA}
\end{figure}

Three different samples were used.

One sample consisted of a flat layer of polyvinylacetate (PVA) several tens of nanometres thick on a \#1 glass coverslip (figure~\ref{fig-PVA}). The coverslip was washed with acetone and etched with a 3:1 solution of sulphuric acid to hydrogen peroxide at 95~\degC\ to remove contaminants. 0.26~g of PVA and 4.33~g of water were mixed to form a 6\% mass/mass PVA solution. The solution was spin-coated on the coverslip at 3,000~rpm for 30~s with 6~s of constant acceleration and deceleration before and after the 30-s constant-speed period. Cuts were made with a sharp razor at different angles in the spin-coated PVA layer in order to create gaps without material. A 120-\textmu m-thick square imaging spacer (Grace BioLabs, OR, USA) with a circular hole 13~mm in diameter was adhered to the coverslip to form a shallow well. The well was filled with water-immersion oil ($n_2=1.3339$, Zeiss Immersol W 2010) and covered with a glass microscope slide. Because the immersion oil reduced the adhesiveness of the imaging spacer, nail varnish was used to create a tight seal around the chamber.

\begin{figure*}
	\begin{center}
		\includegraphics[width=\textwidth]{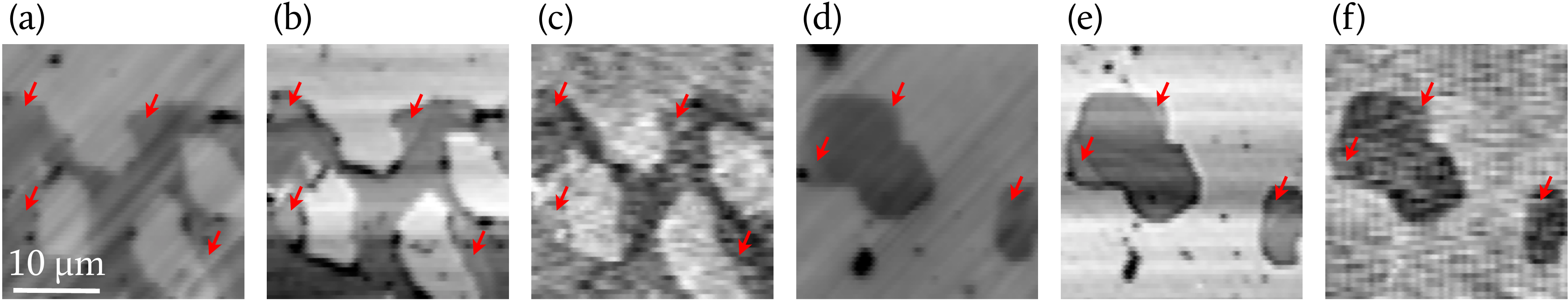}
	\end{center}
	\caption{qDIC phase (a, d), HiRef amplitude (b, e) and HiRef phase (c, f) images of two different fields of view of the ternary-SLB sample. The greyscale values are as follows: a: $m=-0.0456$, $M=-0.0081$; b: $m=0.4371$~V, $M=0.4769$~V; c: $m=-0.1235$, $M=0.1232$; d: $m=-0.0481$, $M=0.0106$; e: $m=0.2497$~V, $M=0.2678$~V; f: $m=-0.1203$, $M=0.1107$. In all images, brighter regions are empty and darker regions are the lipid bilayer. The DOPC domains appear slightly brighter than the sphingomyelin domains in the left and central images; domain boundaries where the change in brightness is visible are indicated by red arrows.}
	\label{fig-bilayer}
\end{figure*}

Each of the other two samples consisted of a supported lipid bilayer (SLB) on a \#1 glass coverslip (figure~\ref{fig-bilayer}). The coverslip was washed and etched as described above. Either pure dipentadecanoylphosphatidylcholine (DC$_{15}$PC) or a ternary lipid solution consisting of dioleoylphosphatidylcholine (DOPC), chicken-egg sphingomyelin and cholesterol in an 11:5:4 molar ratio (DOPC:sphingomyelin:cholesterol) was diluted in isopropanol to a lipid concentration of 1~mg/ml. 150~\textmu l of this solution were deposited on the coverslip and spin-coated as described above; the spin-coating parameters were the same as for the PVA sample. In order to avoid rapid absorption of moisture in the later steps of sample preparation, which would have destroyed the structure of the bilayer, the coverslip was placed in a nitrogen-filled centrifuge tube with a small piece of moist tissue paper and incubated at 37~\degC\ for 1~hr. Finally, an imaging spacer with the aformentioned dimensions was adhered to the coverslip to form a shallow well. The well was filled with degassed phosphate-buffered saline (PBS) and covered with a glass microscope slide. Prior to imaging, the ternary SLB sample was stored at 4~\degC\ for a few hours to allow liquid-ordered (LO) and liquid-disordered (LD) lipid domains to form.

\subsection{Data analysis}\label{subsec-analysis}

The HiRef phase images of the SLB samples were filtered to remove low-frequency noise, which obscured the structural features owing to the low constrast from a single lipid bilayer. This filtering consisted of multiplying the Fourier spectrum of the phase of the image data by $1-f(\nu)$, where $f$ is a sum of gaussian functions, each centred at a frequency $\nu$ and having a frequency width of 1~Hz, to remove frequencies below 5~Hz (phase drift), a collection of frequency peaks near 10~Hz (likely electronic noise from nearby equipment) and 15 harmonics of the fast-axis scan frequency, given by $1/(2N_x t_\text{e})$, where $t_\text{e}=200$~\textmu s is the pixel dwell time and $N_x$ is the number of pixels in one row of the image.

Because the position of the sample stage at any given time during image acquisition is not exactly the commanded position, the position $x'$, $y'$ measured by the stage sensor is recorded at each pixel together with the signal $S'$ at that pixel, forming a time trace of the measurement with the time points $t'$. This time trace is converted into a regular grid of pixels in a post-processing procedure called regularisation. For every pixel position $(x,y)$ on the grid, the signal $S$ is calculated as
\eq{S(x,y) & = & \sum_{t'}S'(t')\,W(t')\left/\sum_{t'}W(t')\right.\,,}
where
\eq{W(t') & = & \exp\left(-\left(\frac{x-x'}{\Delta_x}\right)^2-\left(\frac{y-y'}{\Delta_y}\right)^2\right)}
and the sum is taken over all the time points $t'$ for which $|x-x'|<2\Delta_x$ and $|y-y'|<2\Delta_y$ for computational efficiency. We used $\Delta_x=\Delta_y=\delta_x$. Regularisation improves the quality of the images in terms of accurate representation of the contents of the sample.

After regularisation, line traces with a length of 121 pixels (about $13$~\textmu m) for the PVA sample or 61 pixels (about $7.5$~\textmu m) for the SLB samples were taken along the fast scan axis, the $x$ axis. The positions of the traces were chosen so they were centred along the edge of the PVA or lipid (depending on the sample) and thus roughly half of the pixels on any given trace were on PVA or lipid and the rest were on an empty region; the line traces thus had a step in the middle. The step function
\eq{\Theta(x) & = & \frac{a_1}{2}\,\tanh{\frac{x-a_2}{a_3}}+\frac{a_4}{2}\,\sech{\frac{x-a_5}{a_3}}\nonumber\\
& & +\ a_6(x-a_2)+a_7\vphantom{\frac{b}{b}}}
was simultaneously fitted to the amplitude and phase traces, sharing the spatial parameters ($a_2$, $a_3$ and $a_5$), since features are expected to be at the same locations in the amplitude and phase data. The hyperbolic tangent produces the step, while the hyperbolic secant describes irregularities in the edges, caused for example by the nature of the cuts with the razor (PVA sample) or folds in the lipid (SLB samples). The linear component fits slow drift due to sample tilt and to thermal drifts in the beam paths and in the axial ($z$) position of the sample stage.

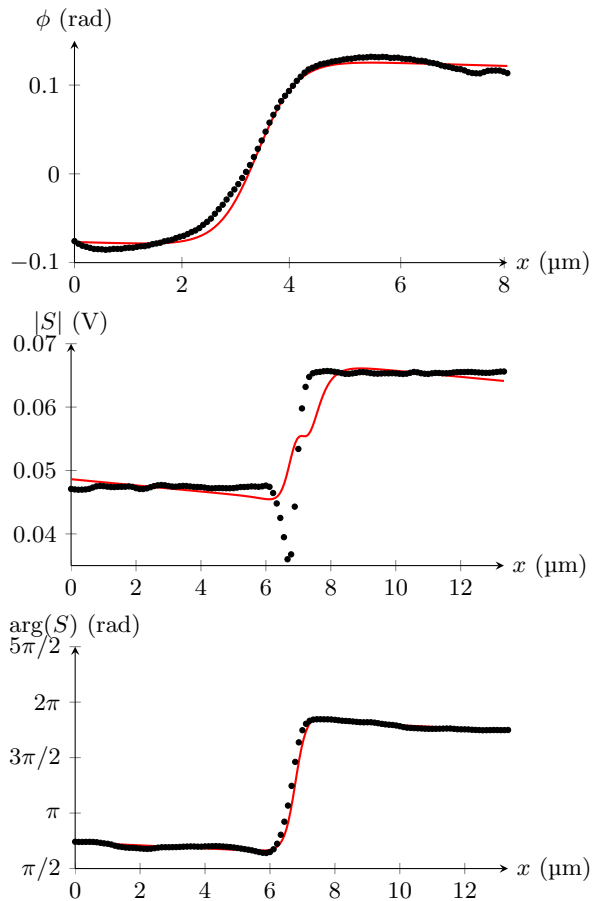
\begin{figure}[t!]
	\begin{tikzpicture}
		\pgfplotsset{every axis legend/.append style={at={(1.5,1)},anchor=north east},samples=100}
		\begin{axis}[width=\graphwidth,height=\graphheight,legend cell align=left,legend style={draw=none,row sep=10pt},xlabel=$x$ (\textmu m),xmin=0,xmax=111,xtick={0,27.55,55.1,83.65,110.2},xticklabels={0,2,4,6,8},ylabel={$\phi$ (rad)},ymin=-0.1,ymax=0.15,scaled ticks=false,yticklabel style={/pgf/number format/.cd,fixed,precision=2},axis x line=bottom,x label style={at={(1,0)},anchor=west},axis y line=left,y label style={at={(current axis.above origin)},anchor=south,rotate=-90}]
		\addplot[only marks,mark size=1pt,black] file[x index=0,y index=1] {lt-qDICeg.dat};]
		\addplot[thick,red,domain=0:111] {(0.2122/2)*tanh(((x-1)-45.91)/9.981)-0.0001255*(x-1)+0.02922};
		\end{axis}
	\end{tikzpicture}\\
	\begin{tikzpicture}
		\pgfplotsset{every axis legend/.append style={at={(1.5,1)},anchor=north east},samples=100}
		\begin{axis}[width=\graphwidth,height=\graphheight,legend cell align=left,legend style={draw=none,row sep=10pt},xlabel=$x$ (\textmu m),xmin=0,xmax=120,xtick={0,18,36,54,72,90,108},xticklabels={0,2,4,6,8,10,12},ylabel={$|S|$ (V)},ymin=0.035,ymax=0.07,scaled ticks=false,yticklabel style={/pgf/number format/.cd,fixed,precision=3},axis x line=bottom,x label style={at={(1,0)},anchor=west},axis y line=left,y label style={at={(current axis.above origin)},anchor=south,rotate=-90}]
		\addplot[only marks,mark size=1pt,black] file[x index=0,y index=1] {lt-reg.dat};
		\addplot[thick,red,domain=0:120,samples=200] {(0.02196/2)*tanh(((x-1)-59.99)/3)-0.00005385*(x-1)+0.05955+(-0.02103/2)/cosh(((x-1)-64.66)/3)};
		\end{axis}
	\end{tikzpicture}\\
	\begin{tikzpicture}
		\pgfplotsset{every axis legend/.append style={at={(1.5,1)},anchor=north east},samples=100}
		\begin{axis}[width=\graphwidth,height=\graphheight,legend cell align=left,legend style={draw=none,row sep=10pt},xlabel=$x$ (\textmu m),xmin=0,xmax=120,xtick={0,18,36,54,72,90,108},xticklabels={0,2,4,6,8,10,12},ylabel={$\arg{S}$ (rad)},ymin=1.5707963268,ymax=7.853981634,ytick={1.570796325,3.14159265,4.7123889804,6.2831853072,7.853981634},yticklabels={$\pi/2$,$\pi$,$3\pi/2$,$2\pi$,$5\pi/2$},scaled ticks=false,yticklabel style={/pgf/number format/.cd,fixed,precision=1},axis x line=bottom,x label style={at={(1,0)},anchor=west},axis y line=left,y label style={at={(current axis.above origin)},anchor=south,rotate=-90}]
		\addplot[only marks,mark size=1pt,black] file[x index=0,y index=1] {lt-peg.dat};]
		\addplot[thick,red,domain=0:120,samples=200] {(3.741/2)*tanh(((x-1)-59.99)/3)-0.004748*(x-1)+4.17924+(0.3549/2)/cosh(((x-1)-64.66)/3)};
		\end{axis}
	\end{tikzpicture}
	\caption{Line traces marked in figure~\ref{fig-PVA}: qDIC phase (top) and HiRef (centre and bottom). Each of the traces is shown with its step function fit $\Theta$. The step function in the qDIC data does not include the hyperbolic secant. Note that the spatial parameters are the same in the two HiRef fits.}
	\label{fig-lt}
\end{figure}

Let us assume for the following discussion that with increasing $x$ the trace moves from the glass-water interface onto the investigated layer. For the phase traces, the constant term $a_7$ represents a phase offset, and the fitted step height $a_1$ is given by
\eq{\arg{\frac{S}{S_{02}}} & = & a_1\label{eq-phi}}
(for comparison with calculations, see also figure~\ref{fig-dn127}). For the amplitude traces, we have
\eq{\left|\frac{S}{S_{02}}\right| & = & \frac{a_7-a_1/2}{a_7+a_1/2}\,,\label{eq-r}}
fully determining $S/S_{02}$. Figure~\ref{fig-lt} shows an example of a line trace from the PVA sample.

The thickness and refractive index were computed numerically from the measured value of $S/S_{02}$ at each line trace. For the computation, the region of $(d,n_1)$ space given by $0\leq d\leq120$~nm and $n_2\leq n_1\leq n_0$ was partitioned finely and the difference between the quantity $S/S_{02}$ calculated with equation~\ref{eq-savg} and the quantity calculated from the line trace fit was minimised; the values of $d$ and $n_1$ assigned to the layer at the line trace were those which minimised this difference.

\section{Results \& discussion}\label{sec-results}

\begin{figure}[b!]
	\begin{center}
		\scalebox{1}{
			\begin{tikzpicture}
			\pgfplotsset{every axis legend/.append style={at={(1.5,1)},anchor=north east}}
			\begin{axis}[xshift=0.52*\graphwidth,yshift=0.11*\graphwidth,width=0.45*\graphwidth,height=0.45*\graphwidth,legend cell align=left,legend style={draw=none,row sep=10pt},xlabel=$|S/S_{02}|$,xmin=0.5,xmax=0.9,ylabel=$\arg{S}$,ymin=-3.14159265,ymax=-1.57079633,scaled ticks=false,ytick={-3.14159265,-2.35619449,-1.57079633,-0.785398163,0},yticklabels={$-\pi$,$-\frac{3\pi}{4}$,$-\frac{\pi}{2}$,$-\pi/4$,$0$},yticklabel style={/pgf/number format/.cd,fixed,precision=2},label style={font=\tiny},tick label style={font=\tiny},y label style={at={(-0.3,0.5)}}]
				\addplot[only marks,red,mark options={scale=0.5}] file[x index=0,y index=1] {PVArp-1.dat};
				\addplot[only marks,orangee,mark options={scale=0.5}] file[x index=0,y index=1] {PVArp-2.dat};
				\addplot[only marks,yelloww,mark options={scale=0.5}] file[x index=0,y index=1] {PVArp-3.dat};
				\addplot[only marks,darkgreen,mark options={scale=0.5}] file[x index=0,y index=1] {PVArp-4.dat};
			\end{axis}
			\begin{axis}[width=\graphwidth,height=\graphwidth,legend cell align=left,legend style={draw=none,row sep=10pt},xlabel=$d$ (nm),xmin=60,xmax=100,ylabel=$n_1$,ymin=1.48,ymax=1.52,scaled ticks=false,ytick={1.48,1.49,1.50,1.51,1.52},yticklabels={1.48,1.49,1.50,1.51,1.52},yticklabel style={/pgf/number format/.cd,fixed,precision=2}]
				\addplot[only marks,red] file[x index=0,y index=1] {PVA-1.dat};
				\addplot[only marks,orangee] file[x index=0,y index=1] {PVA-2.dat};
				\addplot[only marks,yelloww] file[x index=0,y index=1] {PVA-3.dat};
				\addplot[only marks,darkgreen] file[x index=0,y index=1] {PVA-4.dat};
				\addplot[only marks,mark=square,black,error bars/.cd,x dir=both,x explicit,y dir=both,y explicit] coordinates {
(84.7,1.502) +- (6.3,0)};
			\end{axis}
			\end{tikzpicture}}
		\caption{Thickness and refractive index of a PVA film measured simultaneously by HiRef (coloured points). Different colours indicate different fields of view. Uncertainties are $\Delta_d=0.5$~nm and $\Delta_{n_1}=0.0005$. The empty square indicates the result of 472 qDIC measurements assuming a refractive index of 1.502. Inset: Relative amplitude and phase steps measured in PVA. Fit errors (sum of squares) are typically below $10^{-3}$ for $|S/S_{02}|$ and below 0.6~rad$^2$ for $\arg{S}$.}
		\label{fig-resPVA}
	\end{center}
\end{figure}
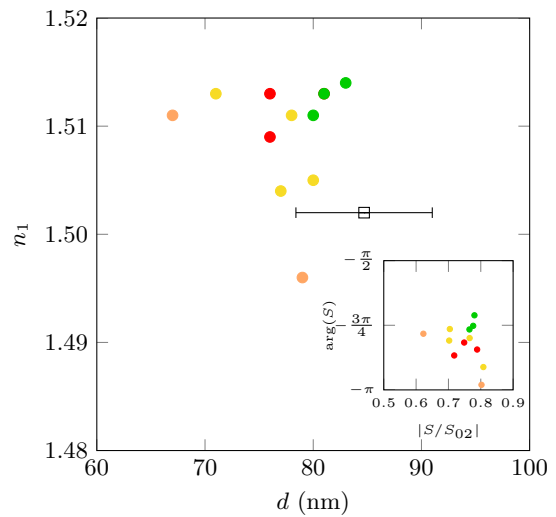

Our HiRef results (12 traces) yield a thickness of $(77.4\pm4.5)$~nm and a refractive index of $1.509\pm0.005$. qDIC measurements (472 traces) yielded a thickness of $(84.7\pm6.3)$~nm assuming a refractive index of $1.502\pm0.017$; this refractive index is the average of a number of literature values.\cite{ref-SchnepfN9,ref-BodurovNN16,ref-MahendiaJAP113,ref-KumarJMO52,ref-Kumar-O119} The uncertainties are the standard deviation in all cases.

Figure~\ref{fig-resPVA} shows the results of the thickness and refractive index measurements. Different colours indicate measurements taken in different fields of view; the empty square indicates the qDIC average for reference. The uncertainty in each individual HiRef measurement, calculated as half of the partition size used for the numerical retrieval of $d$ and $n_1$, are $\Delta_d=0.5$~nm and $\Delta_{n_1}=0.0005$. The standard error of the qDIC measurements is $\Delta_d=0.3$~nm. The inset shows the relative amplitude and phase steps corresponding to the measurements shown in the main graph.

\begin{figure}
	\begin{tikzpicture}
		\pgfplotsset{every axis legend/.append style={at={(1.5,1)},anchor=north east},samples=100}
		\begin{axis}[width=\graphwidth,height=\graphwidth,legend cell align=left,legend style={draw=none,row sep=10pt},xlabel={$\stackrel{\displaystyle d\text{ (nm),}}{\text{HiRef}}$},xmin=40,xmax=110,ylabel={$\stackrel{\displaystyle d\text{ (nm),}}{\text{qDIC}}$},ymin=40,ymax=110,scaled ticks=false,yticklabel style={/pgf/number format/.cd,fixed,precision=0},axis x line=bottom,x label style={at={(current axis.right of origin)},anchor=west},axis y line=left,y label style={at={(current axis.above origin)},anchor=south,rotate=-90}]
		\addplot[only marks,mark size=1pt,black] file[x index=0,y index=1] {corr.dat};
		\end{axis}
	\end{tikzpicture}
	\caption{Comparison of HiRef (horizontal axis) and qDIC (vertical axis) results. Each point corresponds to a pair of measurements, one made with HiRef and one with qDIC, at the same position of a sample.}
	\label{fig-corr}
\end{figure}
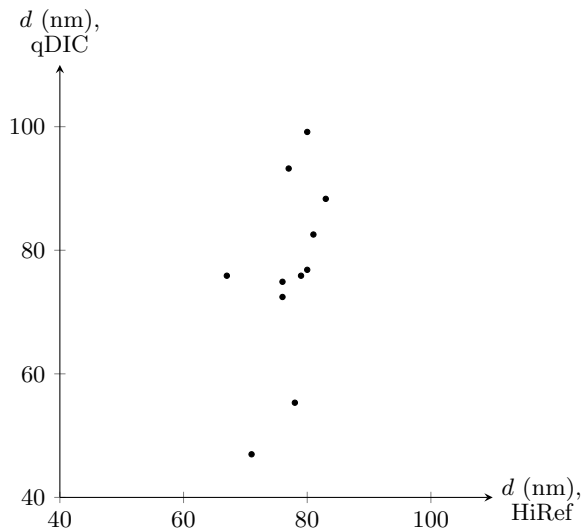

The HiRef measurements are consistent within error with our qDIC measurements. Figure~\ref{fig-corr} shows a comparison of the layer thickness calculated with the two techniques; each point corresponds to one pair of measurements made taking HiRef and qDIC line traces at the same location. The Pearson correlation coefficient is low, only 0.47, indicating that the fluctuations in both techniques arise from different sources. The standard deviation of the HiRef results is only 5.9\% in thickness and 0.3\% in refractive index even with relatively few (12) individual measurements. To estimate the precision of the HiRef measurements, we analysed three traces close to each other, probing virtually the same structure. We found a maximum variation of 3~nm in thickness and 0.007 in refractive index, which is about two thirds (thickness) and half (refractive index) of the uncertainties mentioned above, across the field of view.

The HiRef analysis presented can be used with thin layers of refractive index $n_1$ satsifying $n_2\leq n_1\leq n_0$ if TIR is avoided at any of the interfaces and both the $n_0$ material and the layer in question are transparent. We note, however, that for cases where any of these conditions is not met, equation~\ref{eq-s} can be adapted, taking a different refractive-index range, extinction and/or TIR into account.

Our measurements on the SLB samples explore the limit of the present experimental setup for measuring the thickness and refractive index of a nanometric layer. The HiRef measurements of the thicknesses and refractive indices are $d=(12.7\pm2.9)$~nm and $n_1=1.418\pm0.032$ for DOPC (52 traces) and $d=(9.0\pm3.2)$~nm and $n_1=1.427\pm0.038$ for DC$_{15}$PC (48 traces). While the refractive indices are within error of the literature values of 1.445 for DOPC\cite{ref-DevanathanFEBSJ273} and 1.440 for DC$_{15}$PC,\cite{ref-HowlandBJ92} the thicknesses are two to three times larger than those of single bilayers of the aforementioned lipids (4.1~nm and 5.3~nm, respectively).\cite{ref-ReganL35} However, we show below that the analysis using simulated data with the experimental noise leads to a large systematic error.

Despite the aforementioned inaccuracies in the thickness measurements, it is possible to see sub-nanometre thickness differences as brightness changes in the HiRef images of the ternary-SLB sample (figure~\ref{fig-bilayer}, red arrows). Sphingomyelin enriched with between 60\% and 100\% as much cholesterol as there is sphingomyelin forms bilayers 5.0~nm thick,\cite{ref-ReganL35} which makes the thickness difference between the enriched sphingomyelin domains and the DOPC domains 0.9~nm.

\begin{figure}[b!]
	\begin{center}
		\begin{minipage}[m]{\columnwidth}
			\includegraphics[width=\columnwidth]{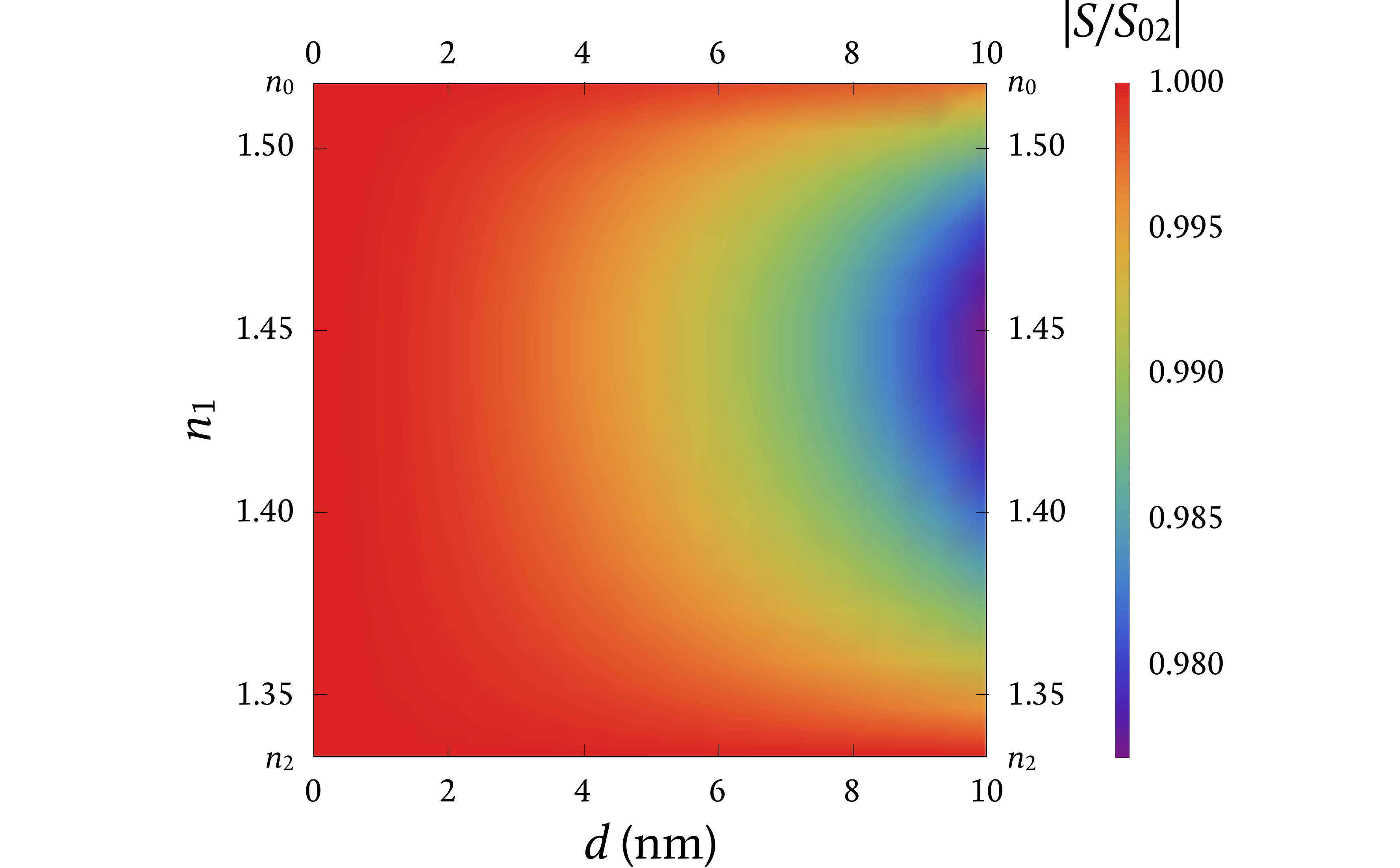}
		\end{minipage}\\[2.5ex]
		\begin{minipage}[m]{\columnwidth}
			\includegraphics[width=\columnwidth]{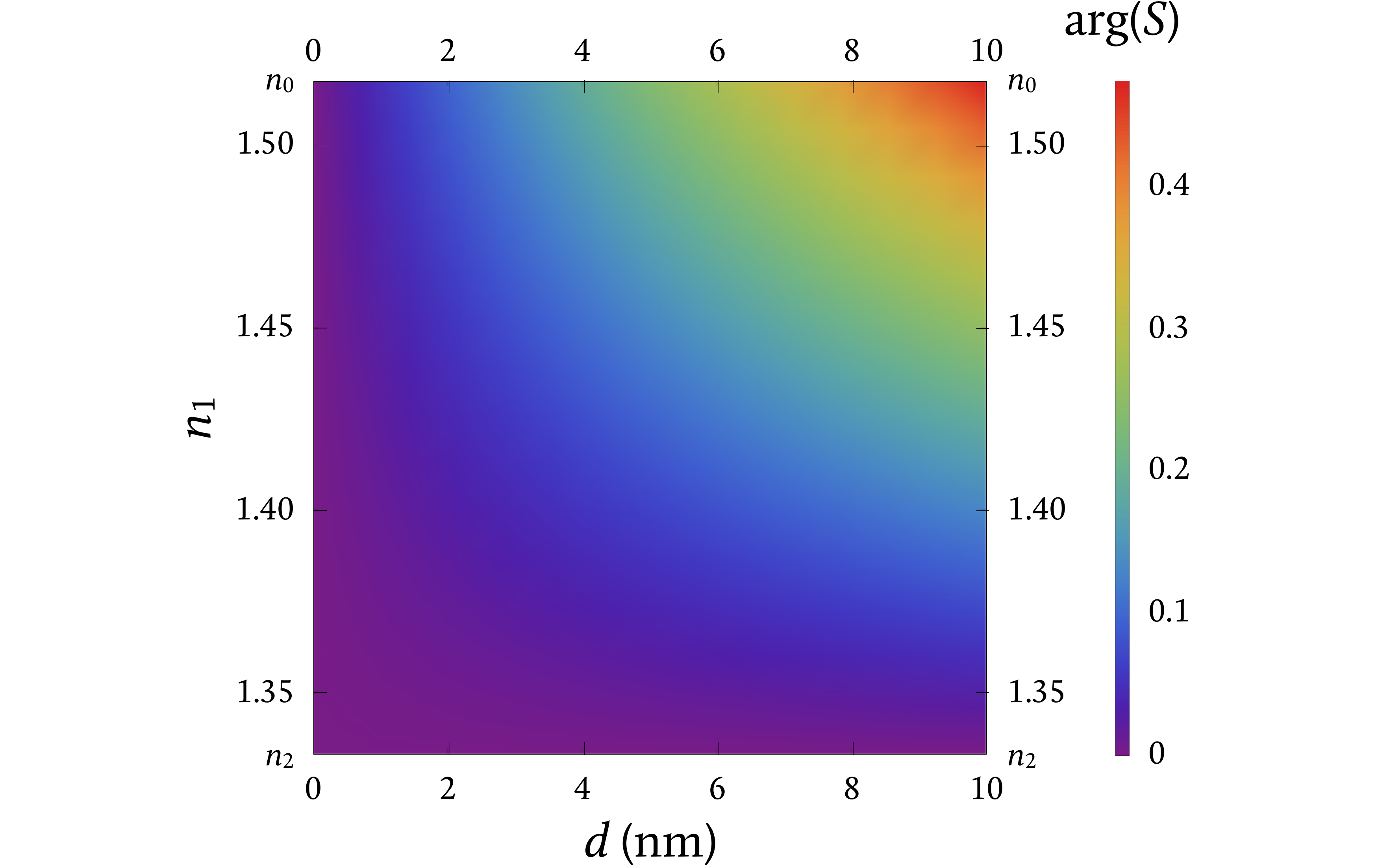}
		\end{minipage}
		\caption{As figure~\ref{fig-s0}, but for a range in $d$ reduced to 10~nm to show the changes in $S/S_{02}$ relevant for single lipid bilayers.}\label{fig-s127red}
	\end{center}
\end{figure}

While HiRef is applicable to very thin layers and sub-nanometre differences in thickness are qualitatively visible in HiRef images of layers 4--5~nm thick (such as cell membranes), accurate thickness and refractive index measurements on such thin layers are hindered by the signal-to-noise ratio of the present data. Figure~\ref{fig-s127red} shows $|S/S_{02}|$ and $\arg{S}$ for an aplanatic objective with a numerical aperture of 1.27 and all parameters as in figure~\ref{fig-s127}, but with a reduced $d$ range suited for the bilayer thicknesses; it can be seen that, for intermediate values of $n_1$ (around 1.45), $|S/S_{02}|$ varies by about 2\% over this range and $\arg{S}$ varies by about 0.2~rad. Notably, the present measurements are not shot-noise-limited; the shot noise is about 53~\textmu V for the chose acquisition parameters. Given that signals are between 250 and 500~mV for the lipid images studied, the shot noise is about 0.01\% of the signal, or 0.05\% of the $\arg{S}$ range and 0.5\% of the $|S/S_{02}|$ range. For the nominal bilayer thickness of 5.0~nm and $n_1=1.45$, this noise results in an uncertainty smaller than 0.5~nm in $d$ and smaller than 0.0005 in $n_1$. This is more than one order of magnitude below the uncertainties determined from the measurements.

\begin{figure}[b!]
	\begin{center}
		\scalebox{1}{
			\begin{tikzpicture}
			\pgfplotsset{every axis legend/.append style={at={(0.55,0.834)},anchor=west}}
			\begin{axis}[width=\graphwidth,height=\graphwidth,legend cell align=left,legend style={},xlabel=$d$ (nm),xmin=0,xmax=12,xtick={0,2,4,6,8,10,12},ylabel=$n_1$,ymin=1.333,ymax=1.518,ytick={1.333,1.35,1.4,1.45,1.5,1.518},yticklabels={$n_2$,1.35,1.40,1.45,1.50,$n_0$},scaled ticks=false,yticklabel style={/pgf/number format/.cd,fixed,precision=2}]
			\addplot[only marks,darkgreen,mark size=1pt] file[x index=0,y index=1] {sim-100.dat};
			\addplot[only marks,yelloww,mark size=1pt] file[x index=0,y index=1] {sim-10.dat};
			\addplot[only marks,red,mark size=1pt] file[x index=0,y index=1] {sim-0.dat};
			\addplot[only marks,mark=square,black,error bars/.cd,x dir=both,x explicit,y dir=both,y explicit] coordinates {
(4,1.44)};
			\legend{$\sigma=5\times10^{-4}$,$\sigma=5\times10^{-5}$,$\sigma=0$}
			\end{axis}
			\end{tikzpicture}}
		\caption{Retrieved thickness $d$ and refractive index $n_1$ from 50 line traces with simulated data with different levels of noise. The empty square indicates the thickness (4~nm) and refractive index (1.440) used in the construction of the data.}
		\label{fig-sim}
	\end{center}
\end{figure}
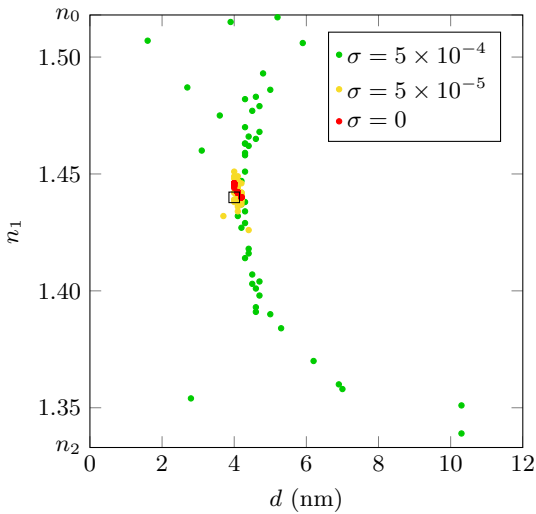

To evaluate the performance of HiRef for the experimental setup if limited by shot noise, we carried out the same data analysis procedure with simulated data. This data consisted of an $N_x\times N_y$ array of complex values with $N_x=N_y=40$ divided into two types of homogeneous regions. The value at one region type was $r_{02}$ for the parameters ($n_0$, $n_2$ and $\lambda$) used for figure~\ref{fig-s0} and normal incidence, and the value at the other was the value of $s$ corresponding to $d=4$~nm and $n_1=1.440$ for the same parameters and normal incidence. Phase noise was then added to the data by generating a $1\times N_x N_y$ array of random numbers with a gaussian distribution with standard deviation $\sigma$ and mean $\mu=0$ (see below). This noise array was then low-pass-filtered, wrapped into an $N_x\times N_y$ array and added to the phase of the simulated data. Versions of the data with $\sigma=0$, $\sigma=5\times10^{-5}$ (which is approximately the shot noise) and $\sigma=5\times10^{-4}$, but otherwise identical to each other, were created and analysed; the version with $\sigma=5\times10^{-4}$ was qualitatively similar to experimental data in terms of the relative strengths of the phase noise and the phase data. The results of this analysis, consisting of 50 line traces in each case, are shown in figure~\ref{fig-sim}. The $(d,n_1)$ values found were $(4.1\pm0.1,1.443\pm0.002)$ with $\sigma=0$, $(4.1\pm0.1,1.442\pm0.005)$ with $\sigma=5\times10^{-5}$, and $(5.2\pm3.482,1.435\pm0.048)$ with $\sigma=5\times10^{-4}$, where $d$ is in~nm. We note that both the error and the uncertainty increase with the phase noise, but even a noise level similar to that seen in the experimental data is not capable of causing an error of 200\% or 300\%, as seen in the experiment. Therefore, the HiRef technique is capable of retrieving (within error) the correct thickness and refractive index.

\section{Conclusions}\label{sec-conc}

We have presented a microscopy technique to measure thickness and refractive index of a thin layer simultaneously and noninvasively. The technique detects the interference between a beam which interacts with the layer and a beam which does not, retrieves the amplitude and phase of the beam reflected by the sample, and numerically computes the thickness and refractive index of the layer from these values using exact expressions for the interference between the beams.

We used a spin-coated PVA layer with a thickness of about 80~nm to test the technique. The measured thickness and refractive index have a standard deviation of only 5.9\% and 0.3\% (respectively) of the measured values and are consistent with qDIC thickness measurements and PVA refractive index values from the literature.

Other microscopy techniques are limited to measuring a single property of the sample (e.g.~qDIC) and/or are invasive (e.g.~fluorescence microscopy, nanoparticle labelling, atomic-force microscopy and electron microscopy). Alternatively, ellipsometry can measure the thickness and the refractive index of a sample at the same time but requires laterally homogeneous samples over large areas and is not spatially resolved at a microscopic scale. Here we have used an optical microscopy technique to measure a sample's thickness and refractive index at the same time in a noninvasive manner.

In the reported experimental data, the results are dominated by classical fluctuations of the amplitude and phase of the detected signal, limiting the technique's ability to reliably measurable layer thicknesses below several tens of nanometres. It is possible to image much smaller thicknesses and changes in thickness and refractive index, such as phase transitions in lipid bilayers, when reducing the classical noise below the shot-noise limit, which for the presented measurements is below 0.5~nm in the thickness and 0.0005 in the refractive index, more than an order of magnitude below the present classical noise. Therefore, a suited referencing technique is required, which could consist of a common-path referencing using a defocussed beam with a different heterodyne frequency shift. This would allow us to address important questions which are inaccessible with current techniques. For example, the difference in thickness between lipid domains in different thermodynamic phases is about 0.9~nm.\cite{ref-ReganL35} Such measurements, performed on live neurons rather than artificial lipid bilayers, might then be able to probe whether or not neural activity involves a phase transition,\cite{ref-NahmadRohenPhDthesis} as proposed by Heimburg and Jackson in 2005.\cite{ref-HeimburgPNAS102} Another potential application would be the observation of lipid rafts in cell membranes.\cite{ref-MunroC115}

The data presented in this work is available from the Cardiff University data archive, http://doi.org/10.17035/xxx.

\section{Acknowledgements}

A.N.R.\ acknowledges support of his PhD studies by CONACYT (scholarship recipient number 581516). The microscope setup used was developed within the UK EPSRC Leadership fellowship award of P.B.\ (grant number EP/I005072/1 and equipment grant number EP/M028313/1). The authors thank Iestyn Pope for support in instrument development and data acquisition and George Zorinyants and Joseph Williams for their contributions to the qDIC data analysis.

\section{References}

\begingroup
\renewcommand{\section}[2]{}

\endgroup

\vfill

\end{document}